\documentclass[12pt,preprint]{aastex}
\usepackage{amsbsy}
\usepackage{amsmath}
\newcommand{\be}{\begin{equation}}
\newcommand{\ee}{\end{equation}}

\def\ltsima{$\; \buildrel < \over \sim \;$}

\def\lsim{\lower.5ex\hbox{\ltsima}}
\def\gtsima{$\; \buildrel > \over \sim \;$}
\def\gsim{\lower.5ex\hbox{\gtsima}}
\shorttitle{Superfluid vortex unpinning as a coherent noise process}
\shortauthors{Melatos \& Warszawski}
\begin{document}
\title{Superfluid vortex unpinning as a coherent noise process,
 and the scale invariance of pulsar glitches}

\author{A. Melatos\altaffilmark{1} and L. Warszawski\altaffilmark{1}}

\email{a.melatos@physics.unimelb.edu.au}

\altaffiltext{1}{School of Physics, University of Melbourne,
Parkville, VIC 3010, Australia}

\begin{abstract}
\noindent 
The scale-invariant glitch statistics observed in individual pulsars (exponential waiting-time and power-law size distributions) are consistent with a critical self-organization process, wherein superfluid vortices pin metastably in macroscopic domains and unpin collectively via nearest-neighbor avalanches. Macroscopic inhomogeneity emerges naturally if pinning occurs at crustal faults. If, instead, pinning occurs at lattice sites and defects, which are macroscopically homogeneous, we show that an alternative, noncritical self-organization process operates,  termed {\em coherent noise}, wherein the global Magnus force  acts uniformly on vortices trapped in a range of pinning potentials and undergoing thermal creep. It is found that vortices again unpin collectively, but not via nearest-neighbor avalanches, and that, counterintuitively, the resulting glitch sizes are scale invariant, in accord with observational data. A mean-field analytic theory of the coherent noise process, supported by Monte-Carlo simulations, yields a power-law size distribution, between the smallest and largest glitch, with exponent $a$ in the range $-2\leq a \leq 0$. When the theory is fitted to data from the nine most active pulsars, including the two quasiperiodic glitchers  PSR J0537$-$6910 and PSR J0835$-$4510, it directly constrains the distribution of pinning potentials in the star, leading to two conclusions: (i) the potentials are broadly distributed,  with the mean comparable to the standard deviation; and (ii) the mean potential decreases with characteristic age. Fitting the theory to the data also constrains the pinned vortex fraction and the rate of thermal creep. An observational test is proposed to discriminate between nearest-neighbor avalanches and coherent noise: the latter process predicts a statistical excess of large glitches (`aftershocks') following a large glitch, whereas the former process does not.  Its discriminatory power is discussed under various microphysical scenarios.
\end{abstract}


\section{Introduction 
 \label{sec:coh1}}
The number of pulsar glitches recorded historically has quadrupled
in the last five years and now approaches 300 events,
following improvements in the sensitivity and duty cycle of
radio pulsar timing, pioneered by the Parkes Multibeam Survey
\citep{hob02,kra03,man05,jan06,mid06,per07}.
It is now possible to disaggregate
the data and measure reliably the glitch size and waiting-time
distributions in individual pulsars.
\citet{mel08} showed that the distributions are consistent
with power laws and exponentials respectively in nine objects
(two of which also feature a smaller quasiperiodic component),
suggesting that the glitch mechanism is predominantly
scale invariant and obeys Poisson statistics
\citep{won01,alp06,mel08}.

Scale invariance and Poisson statistics are universal characteristics
of self-organized critical systems,
in which discrete, interacting elements adjust
in response to a slow, localized, external driver
via intermittent avalanches, i.e.\ 
nearest-neighbor ``domino chains''
of local, impulsive, threshold-activated relaxation events
\citep{jen98,sor04}.
Self-organized criticality is observed widely in nature,
for example in sandpiles \citep{bak87}, earthquakes \citep{sor04}, 
solar flares \citep{lu91,whe00},
and magnetized type II superconductors
\citep{fie95}.
Two popular pulsar glitch paradigms, 
involving crust fracture
\citep{alp96,mid06}
and collective unpinning of superfluid vortices 
\citep{and75,che88,alp96},
lead naturally to avalanche dynamics \citep{mel08}
and can be simulated efficiently with cellular automata
\citep{mor96,war08}.

As avalanches traverse a self-organized critical system, 
they leave behind inhomogeneities on all scales, 
up to and including the system size.
In the vortex unpinning paradigm for pulsar glitches, 
pinned vortices cluster in metastable reservoirs surrounded by depletion zones,
i.e.\ `capacitive elements' 
\citep{che88,alp96,won01}.
Some reservoirs occupy a sizable portion of the star.
This picture works admirably if pinning occurs along a macroscopic network
of crustal faults (created by seismic disturbances, for example)
\citep{mid06}.
It is harder to reconcile with pinning at nuclear lattice defects (e.g.\ interstitial vacancies) or simply lattice sites,
which are separated by tens of lattice spacings and are therefore 
homogeneous on macroscopic scales
\citep{jon98,don03,avo07}.
In this picture vortices hop between adjacent
pinning sites in response to thermal fluctuations and the Magnus stress
(from differential spin down of the crust and superfluid)
without deviating macroscopically from a regular Abrikosov lattice
\citep{lin93}.

A central puzzle concerning vortex unpinning is how it accounts for the scale invariance of pulsar glitches. Normally, scale invariance is a sign of collective behavior, involving correlations between inhomogeneities on all scales, including the largest. But, as explained above, large-scale inhomogeneity is hard to contrive when pinning occurs at lattice defects or lattice sites, so the inter-vortex forces approximately balance.  How, then, do vortices unpin in sympathy in such large (macroscopic) numbers? Furthermore, why do glitch sizes vary by many orders of magnitude (four decades in one object) from glitch to glitch?  We emphasize at this point that the size of a glitch is related to not only the number of vortices that unpin but also the distance moved in the radial direction before repinning \citep{alp84,che88,jah06}.  If nuclear lattice pinning is characterized by a typical pinning potential,or even a moderate range of pinning potentials, as theory predicts \citep{jon98,don06,avo07}, and if the global Magnus stress is felt uniformly by all pinned vortices, one naively expects glitches in any individual pulsar to recur periodically (whenever the Magnus stress rises to match the typical pinning force) and hence display approximately equal sizes, contrary to what is seen. These are profound challenges for all glitch mechanisms. 

In this paper, we invoke the {\em coherent noise} process introduced by \citet{sne97} to describe the collective dynamics of vortex unpinning and calculate the resulting glitch statistics. \footnote{The term `coherent noise' is something of a misnomer when applied to vortex unpinning.  The noisy element, namely thermal creep, does not operate coherently throughout the star; rather, the vortices feel a globally coherent Magnus stress. Arguably, the adjective `coherent stress' describes the model better. However, we elect to retain the original terminology in this paper in order to preserve consistency with the statistical mechanics literature.} Remarkably, we find that scale invariance emerges automatically between the minimum and maximum sizes, even without nearest-neighbor avalanches and large-scale inhomogeneity. In \S\ref{sec:coh2}, we specify the model and calculate the glitch size distribution analytically from first principles, in the stationary, mean-field approximation. The results are expressed in terms of three principal variables: the mean glitch rate, which is observable; the mean thermal unpinning rate, which does not equal the mean glitch rate; and the distribution of pinning potentials, which can be predicted from nuclear physics. In \S\ref{sec:coh3}, we compare the analytic theory against the output of Monte-Carlo simulations and show how existing and future observational data can be used to constrain the distribution of pinning potentials. By way of illustration, the theory is fitted to data from the nine  most active glitchers currently known. In \S\ref{sec:coh4}, we explore the implications of our results for the nuclear microphysics of vortex pinning. We also propose an observational test that may discriminate between coherent noise and nearest-neighbor avalanches.

We emphasize, at the outset, that the coherent noise mechanism does not operate under
conditions when pinning does not occur
\citep{jon97,jon98}.

\section{Coherent noise mechanism
 \label{sec:coh2}}
\citet{sne97} first postulated the coherent noise mechanism to describe 
discrete, far-from-equilibrium systems driven by a global stress,
which acts on all elements in the system simultaneously.
Elements respond by relaxing locally via threshold-activated events.
Coherent noise offers a pathway to self-organization which is
neither critical nor interaction-dominated;
for instance, the stationary state is homogeneous on average,
without long-range spatial correlations mediated by nearest-neighbor interactions.
By contrast, in critical systems like sandpiles,
the external driver is localized, and its influence propagates
across the system over time via scale-invariant avalanches
\citep{bak87,jen98}.
Nevertheless, counterintuitively, 
coherent noise does produce intermittent, collective events 
with a scale-invariant size distribution,
just like self-organized critical systems.
\footnote{
A word of caution: self-organized criticality and coherent noise 
are not the only collective mechanisms that lead to intermittency
and scale invariance in far-from-equilibrium systems.
Other examples include percolation, multiplicative noise,
and highly optimized tolerance
\citep{sor04}.
}

In this section, we describe a simple cellular automaton that models
the collective dynamics of vortex unpinning in a neutron star as a
coherent noise process (\S\ref{sec:coh2a} and \S\ref{sec:coh2b}).
We then solve analytically for the mean-field behavior of the model
(\S\ref{sec:coh2c})
and derive the distribution of glitch sizes
as a function of the pinning parameters (\S\ref{sec:coh2d}).

\subsection{Pinning at lattice sites and defects
\label{sec:coh2a}}
Consider a neutron star containing $N$ pinned superfluid vortices, each carrying circulation $\kappa$, amounting to a fraction $\epsilon$ of all the vortices in the star. In this paper, we take $N$ to be constant, as we seek to model the short-term glitching of individual pulsars, monitored over decades, rather than the long-term glitching of the whole pulsar population,  which evolves on the spin-down time-scale $\nu/\dot{\nu}$. Here, $\nu$ is the spin frequency.

The microscopic cause of pinning, whether it be the attractive force of the lattice nuclei themselves, defects in the lattice, or macroscopic faults in the stellar crust, is an important factor in determining the degree of homogeneity of pinning sites.  Pinning at lattice nuclei, the default assumption in this paper, is the most homogeneous.  Interstitial pinning at the midpoint between nuclei \citep[if it exists; see][]{avo07} is the next most homogeneous level of pinning. Actual defects (e.g. nuclear impurities like monovacancies or shear layers) are estimated to occur roughly every $\sim 30$ nuclei \citep{deb98}, but these too are essentially ``perfectly homogeneous'' on the vortex lattice scale that matters in this paper. Macroscopic faults in the solid crust, e.g. ``tectonic" plates formed by large-scale cracking \citep[if it exists; see][]{mid06} are probably inhomogeneous on the vortex lattice scale and lie outside the scope of any coherent noise model. Suppose each pinned vortex (labelled $i$, with $1\leq i \leq N$)  occupies the site of a defect  in the nuclear lattice \citep{deb98}, whose pinning threshold  (expressed as a force per unit length) is denoted by $F_{\rm p}^{(i)}$. If the Magnus force $F_{\rm M}$ exceeds $F_{\rm p}^{(i)}$, then the vortex unpins and moves with the local superfluid flow. However, pinning sites are abundant microscopically, occurring once every $\sim 30$ lattice spacings  \citep{jon98,lin02}.  Consequently, the unpinned vortex moves at most $\sim 10^{-13}\,{\rm m}$, much less than the mean vortex separation $(\kappa/4\pi\nu)^{1/2}$, before it immediately repins at a new defect, with a new pinning threshold $F_{\rm p}^{(j)}$ [$\neq F_{\rm p}^{(i)}$ in general]. The microscopic abundance of pinning sites is the key difference between this scenario and one driven by crust fracture, to which avalanche models apply; at all times, the pinned vortices form something close to a regular Abrikosov array,  which, like the underlying pinning sites, is homogeneous on large scales. \footnote{In this paper, we declare the vortices to be rectilinear, although recent work suggests that the meridional flow induced by crust-superfluid differential rotation excites the  Glaberson-Donnelly instability and generates a vortex tangle \citep{per05,per06,mel07}.} 

Let $\phi(F_{\rm p}) dF_{\rm p}$ be the fraction of defects whose
pinning threshold lies between $F_{\rm p}$ and $F_{\rm p}+dF_{\rm p}$.
In principle, it is possible to predict $\phi(F_{\rm p})$ 
{\em ab initio} from nuclear structure calculations,
although attempts to do so have yielded conflicting results due to
the subtlety of the physics.
One approach, based on the local density approximation,
suggests that pinning in the inner crust is strongest at intermediate densities
and is predominantly interstitial \citep{don03,don04} or nuclear \citep{don06}.
It yields pinning energies $E_{\rm p}$ in the range 
$1\,{\rm MeV} \lesssim E_{\rm p} \lesssim 4\,{\rm MeV}$,
although $E_{\rm p}$ can be $\sim 20$ times greater if the normal
component of the superfluid is absent (``pure phase'').
A second approach, based on the mean-field, Hartree-Fock-Bogoliubov approximation,
suggests that pinning is strongest 
at low ($5\times 10^{15}\,{\rm kg\,m^{-3}}$)
and high ($2\times 10^{17}\,{\rm kg\,m^{-3}}$) densities
\citep{lin91,avo07}.
It yields
$1\,{\rm MeV} \lesssim E_{\rm p} \lesssim 3\,{\rm MeV}$
for Fermi momenta in the range 0.5--1$\,{\rm fm^{-1}}$.
Yet another line of argument suggests that pinning is too weak to occur
at all, unless the concentration of monovacancies is unexpectedly high
\citep{jon97,jon98}.
Summarizing the foregoing results,
we adopt as a crude working hypothesis the top-hat distribution
\begin{equation}
 \phi(F_{\rm p})
 =
 (2\Delta)^{-1} 
 H(F_{\rm p} - F_0 + \Delta)
 H(-F_{\rm p} + F_0 + \Delta)~,
 \label{eq:coh1}
\end{equation}
where $F_0$ is the mean,
$\Delta$ is the half-width,
and $H(\dots)$ is the Heaviside step function.
(Note that we assume $F_{\rm p}\geq 0$ in this paper.)
Typical values are
$F_0 = 1\times 10^{16}\,{\rm N\,m^{-1}}$ and
$\Delta = 6\times 10^{15}\,{\rm N\,m^{-1}}$
\citep{don06,avo07},
taking the superfluid coherence length to be $\xi=5\,{\rm fm}$.

Remarkably, any physically reasonable form of $\phi(F_{\rm p})$
(e.g.\ top hat, power law, Gaussian)
generates a scale invariant distribution of event sizes
over some interval
\citep{sne97}.
We explain why in \S\ref{sec:coh3}.
This robustness is an attractive feature of the coherent noise mechanism
as a model for pulsar glitches.
Note that (\ref{eq:coh1}), as written, is independent of position
within the star.
We make this simplification here in anticipation of the 
mean-field analysis to be carried out in \S\ref{sec:coh2c},
but of course it is not realistic;
pinning theories favor certain ranges of superfluid density
\citep{don03,don04,don06,avo07}.
We will generalize (\ref{eq:coh1}) by letting it vary with position
in a forthcoming paper.

\subsection{Cellular automaton for vortex unpinning
 \label{sec:coh2b}}
Let us now evolve the pinned vortices in discrete time steps $\Delta t$,
with the aid of a simple cellular automaton. 
At each time step,
the following four rules are applied to update the state of the automaton.
They encode the microphysics of vortex unpinning in an idealized fashion,
but the resulting {\em collective} behavior is insensitive to the 
details of the microphysics, as much accumulated experience with
cellular automata shows \citep{jen98,sor04}.
\begin{enumerate}
\item
A value of the global stress (here, the Magnus force) $F_{\rm M}$
is chosen at random from a probability distribution function $\psi(F_{\rm M}$).
As the Magnus force originates from crust-superfluid differential spin down,
$\psi(F_{\rm M})$ shares the same form as the observed glitch waiting-time
distribution (see below).
\item
A small fraction $f\ll 1$ of the $N$ pinned vortices unpin at random,
e.g. due to thermal fluctuations. 
For simplicity, we do not bias this random process towards particular 
pinning sites.
A more realistic model might preferentially unpin those vortices 
with $F_{\rm p}^{(i)}$ just above $F_{\rm M}$,
e.g.\ with $F_{\rm M} \leq F_{\rm p}^{(i)} \leq F_{\rm M} + k_{\rm B}T/ \xi^2$,
where $T$ is the temperature of the crust and $k_{\rm B}$ is
Boltzmann's constant.
\item
The stress $F_{\rm M}$ is applied simultaneously to all the 
remaining $(1-f)N$ pinned vortices,
viz.\ $F_{\rm M}^{(i)} = F_{\rm M}$.
All vortices with $F_{\rm p}^{(i)} < F_{\rm M}$ unpin.
For simplicity, we take $F_{\rm M}$ to be uniform, but it is straightforward
to let it vary realistically with distance from the rotation axis.
Generalizing the model in this way does not alter its collective
behavior at all.
\item
Each unpinned vortex repins almost immediately at a nearby defect
(see \S\ref{sec:coh2a}) and is assigned a new threshold.
\end{enumerate}

In the simplest version of the automaton, 
the pinned vortices do not interact,
except implicitly through the mutual repulsion which keeps them in a
regular Abrikosov array. 
Consequently, in the stationary state,
long-range spatial correlations do not emerge,
and the occupied pinning sites are distributed homogeneously
(see \S\ref{sec:coh2}; transitory correlations can arise by accident, of course).
Macroscopic homogeneity is preserved even
if nearest-neighbor vortex interactions are allowed,
provided that the interactions are weaker than the global stress
most of the time.
On the other hand, in the interaction-dominated regime,
homogeneity breaks down,
and the coherent noise process transitions to an avalanche process
\citep{war08}.

What is the form of $\psi(F_{\rm M})$?  There is strong empirical evidence that glitch waiting times obey Poisson statistics, e.g.\ from a Kolmogorov-Smirnov analysis of nine individual pulsars \citep{won01,mel08,war08} and the interpretation of anomalous braking indices in terms of unresolved glitches \citep{joh99,alp06}.  Hence the automaton is asynchronous; $\Delta t$, the time since the last glitch, is different at every time step and exponentially distributed.  A physically reasonable, monotonically decreasing form of $\psi(F_{\rm M})$ is therefore the Poisson-like distribution
\begin{equation}
 \psi(F_{\rm M}) = \sigma^{-1} \exp(-F_{\rm M}/\sigma)~,
 \label{eq:coh2}
\end{equation}
where $\sigma$ is a characteristic value of the Magnus stress that accumulates prior to a glitch.  Possible definitions of $\sigma$ in terms of the physical parameters of neutron stars are discussed in \S\,\ref{sec:coh4a}.  For the remainder of this section and \S\,\ref{sec:coh3}, $\sigma$ is treated as a stress scale factor when analyzing the behaviour of the model, which depends on $\sigma$ only through the dimensionless combinations $F_0/\sigma$ and $\Delta /\sigma$.  Crucially, the collective dynamics are the same irrespective of the exact form of $\psi(F_{\rm M})$.  \citet{sne97} showed that any distribution that falls off sufficiently rapidly at large $F_{\rm M}$, such that $\int_{F_{\rm M}}^\infty dx\, \psi(x)$ scales as $[\psi(F_{\rm M})]^\alpha$ ($\alpha$ real) to leading order, generates a power-law distribution of event sizes over some size interval.

In the absence of nearest-neighbor interactions, one must have $f\neq 0$.  Otherwise, the system stagnates ultimately, with $\langle F_{\rm M} \rangle \ll F_{\rm p}^{(i)}$ for all $i$ and vortices unpinning ever more rarely as time passes. The competition between coherent forcing and thermal creep is  therefore essential for the system to develop scale invariance. In more elaborate models incorporating nearest-neighbor interactions, thermal creep is strictly unnecessary and one can set $f=0$, but, as discussed in \S\ref{sec:coh1}, this is not believed to be the situation in pulsars \citep{lin93,lin96}. Note that $f$ must be small to account for a power-law distribution of event sizes over several decades \citep{sne97}.

\subsection{Mean-field master equation
 \label{sec:coh2c}}
We now analyse the cellular automaton
in the mean-field approximation, exploiting the property of homogeneity.
Let $g(F_{\rm p}) dF_{\rm p}$ be the time-averaged fraction of
pinned vortices trapped in pinning sites with thresholds in the range
$F_{\rm p} \leq F_{\rm p}^{(i)} \leq F_{\rm p}+dF_{\rm p}$.
Crucially, $g(F_{\rm p})$ is not the same as $\phi(F_{\rm p})$;
the latter function counts all pinning sites without bias,
whereas $g(F_{\rm p})$ counts only those sites that
are actually occupied in the stationary state.
To illustrate this key point, consider two sites $i$ and $j$,
with $F_{\rm p}^{(i)} = F_0 -\Delta/2$ and $F_{\rm p}^{(j)} = F_0 +\Delta/2$.
If $\phi(F_{\rm p})$ is given by (\ref{eq:coh1}), say, then we have
$\phi[F_{\rm p}^{(i)}] = \phi[F_{\rm p}^{(j)}]$,
yet we expect
$g[F_{\rm p}^{(i)}] < g[F_{\rm p}^{(j)}]$,
as it is easier to unpin from the shallower potential $i$.

Now consider how the number of pinned vortices with
thresholds in the range
$F_{\rm p} \leq F_{\rm p}^{(i)} \leq F_{\rm p}+dF_{\rm p}$
changes during one time step.
There are $Ng(F_{\rm p})dF_{\rm p}$ such vortices at the start of
the time step.
According to rule two in \S\ref{sec:coh2b},
$Nfg(F_{\rm p})dF_{\rm p}$ vortices unpin from this threshold range
(or indeed any other threshold range of width $dF_{\rm p}$,
as rule two is unbiased with respect to $F_{\rm p}$ in this paper)
due to thermal creep.
According to rule three,
all the remaining $N(1-f)g(F_{\rm p})dF_{\rm p}$ vortices unpin
if $F_{\rm p}$ is less than $F_{\rm M}$,
an eventuality which occurs with probability
$\int_{F_{\rm p}}^\infty dF_{\rm M}\, \psi(F_{\rm M})$;
otherwise, none unpin.
Finally, according to rule four,
a number of vortices repin at sites in the threshold range
$[F_{\rm p},F_{\rm p}+dF_{\rm p}]$.
The number is clearly proportional to $N\phi(F_{\rm p})dF_{\rm p}$,
as $\phi(F_{\rm p})$ is the threshold distribution
presented to an unpinned vortex by the nuclear lattice as it is about
to repin.
The constant of proportionality $A$ is determined by normalization.
(Recall that $N$ is constant in this paper, as we are interested
in the glitch dynamics over decades.)
In summary, we can write down the following
master equation for $g(F_{\rm p})$:
\begin{eqnarray}
 & & 
 \Delta t \frac{\partial[Ng(F_{\rm p}) dF_{\rm p}]}{\partial t} 
 \nonumber \\
 & & 
 =
 A N \phi(F_{\rm p})dF_{\rm p}
 - Nfg(F_{\rm p})dF_{\rm p}
 \nonumber \\
 & &  
 \phantom{=}
 - N(1-f)g(F_{\rm p})dF_{\rm p}
   \int_{F_{\rm p}}^\infty dF_{\rm M}\, \psi(F_{\rm M})~.
\label{eq:coh4}
\end{eqnarray}

In the stationary state, the left-hand side of (\ref{eq:coh4}) vanishes, 
and we obtain
\begin{equation}
 g(F_{\rm p})
 =
 A \phi(F_{\rm p})
  \left[ f + (1-f) \int_{F_{\rm p}}^\infty dF_{\rm M}\, \psi(F_{\rm M}) \right]^{-1} ,
\label{eq:coh5}
\end{equation}
with $A$ fixed by the normalization condition
$\int_0^\infty dF_{\rm p} \, g(F_{\rm p}) = 1$.
An explicit analytic formula for $g(F_{\rm p})$ is presented in
equation (\ref{eq:cohappa1}) of Appendix \ref{sec:cohappa}
for the particular choices of $\phi(F_{\rm p})$ and $\psi(F_{\rm M})$
given by (\ref{eq:coh1}) and (\ref{eq:coh2}) respectively.
With these choices,
$g(F_{\rm p})$ is zero outside the interval
$|F_{\rm p} - F_0| \leq \Delta$,
where no pinning sites are available.

\subsection{Glitch sizes
 \label{sec:coh2d}}
The time-averaged probability distribution function of glitch sizes $s$, denoted by $h(s)$, follows directly from $g(F_{\rm p})$. Let $s=\Delta\nu/\nu$ be the fractional rise in spin frequency during a glitch, as measured in pulsar timing experiments. For simplicity, we take $s$ to be the number of vortices that unpin during a time step, divided by the total number of vortices (pinned and unpinned) in the star. [In reality, the contribution to $\Delta\nu/\nu$ from each vortex is also proportional to the distance it moves radially before repinning; see \S\ref{sec:coh1},  rule three in \S\ref{sec:coh2b}, the worked example in \S,\ref{sec:obs} and \citet{alp96}.] If the global Magnus stress applied at that time step is $F_{\rm M}$, then the resulting glitch size is given by 
\begin{equation}
 s(F_{\rm M})
 =
 \epsilon f + \epsilon (1-f) \int_0^{F_{\rm M}} dF_{\rm p} \, g(F_{\rm p})~.
\label{eq:coh6}
\end{equation}
The first term on the right-hand side of (\ref{eq:coh6})
was omitted by \citet{sne97}; it is small ($f\ll 1$).
An analytic formula for $s(F_{\rm M})$ is presented in
equation (\ref{eq:cohappa2}) of Appendix \ref{sec:cohappa}.
With $\phi(F_{\rm p})$ chosen according to (\ref{eq:coh1}),
one obtains $s = \epsilon f$ for all $F_{\rm M} < F_0-\Delta$
and $s = \epsilon$ for all $F_{\rm M} > F_0+\Delta$.

Equation (\ref{eq:coh6}) defines $F_{\rm M}$ implicitly (and uniquely)
as a function of $s$
in the interval $|F_{\rm M} - F_0| \leq \Delta$.
Hence the probability of getting a glitch with size in the range $[s,s+ds]$
equals the probability that the global Magnus stress lies in the interval
$[F_{\rm M}(s),F_{\rm M}(s)+dF_{\rm M}]$,
i.e.\ $\psi(F_{\rm M})dF_{\rm M}$,
which transforms into
$\psi[F_{\rm M}(s)] [dF_{\rm M}(s)/ds] ds$ after changing variables
by applying the chain rule. Combining this result with (\ref{eq:coh6}),
we arrive at
\begin{equation}
 h(s) = \frac{\psi[F_{\rm M}(s)]}{\epsilon(1-f) g[F_{\rm M}(s)]}~.
\label{eq:coh7}
\end{equation}
Equation (\ref{eq:coh7}) can be evaluated formally by inverting (\ref{eq:coh6})
to obtain $F_{\rm M}(s)$ or, more practically, by evaluating 
$s(F_{\rm M})$ from (\ref{eq:coh6}),
$h(F_{\rm M})$ from (\ref{eq:coh7}),
and then graphing $h(s)$ parametrically.
An analytic formula for $h(s)$ is presented in
equation (\ref{eq:cohappa3}) of Appendix \ref{sec:cohappa}.
Care must be exercised
wherever $dF_{\rm M}(s)/ds$ diverges and $s(F_{\rm M})$ is not invertible.
For example, 
with $\phi(F_{\rm p})$ chosen according to (\ref{eq:coh1}),
$h(s)$ is bracketed by two delta-function spikes,
$( 1 - e^{-F_0+\Delta} ) \delta(s - \epsilon f)$
and 
$e^{-F_0-\Delta} \delta(s-\epsilon)$, 
corresponding to events with
$F_{\rm M} \leq F_0-\Delta$ (thermal creep only)
and
$F_{\rm M} \geq F_0+\Delta$ (all vortices unpin)
respectively.
Even if $\phi(F_{\rm p})$, and hence $g(F_{\rm p})$, are nonzero for all $F_{\rm M}$
[e.g.\ if $\phi(F_{\rm p})$ is Gaussian],
the spikes remain:
the delta functions are smeared out somewhat,
but $h(s)$ still diverges steeply (and integrably)
as $s \rightarrow \epsilon f$ and $s\rightarrow \epsilon$.
It is easy to verify this result analytically or with Monte-Carlo simulations.

\section{Glitch statistics
 \label{sec:coh3}}
The glitch size distribution $h(s)$ is a power law over many decades for $f\ll 1$ \citep{sne97}.  Importantly, there are maximum and minimum cut-offs to the glitch size, which set natural limits on the extent of the power law.  Its shape and normalization contain information about the pinning parameters in $\phi(F_{\rm p})$.  In this section, we demonstrate that it is possible in practice to infer these parameters by fitting the theory to observational data.  In \S\ref{sec:coh3a} and \S\ref{sec:coh3b},
we implement the four rules in \S\ref{sec:coh2b} in a simple
Monte-Carlo simulation and study systematically how $h(s)$ varies
as a function of $F_0/\sigma$ and $\Delta/\sigma$.
The simulation results are independent of $N$ for $Nf \gg 1$.
In \S\ref{sec:coh3c} and \S\ref{sec:coh3d},
we fit the theoretical form of $h(s)$ to data from the nine most active glitchers,
both Poissonian and quasiperiodic,
in order to derive constraints
on $F_0/\sigma$, $\Delta/\sigma$, $\epsilon$, and $f$.

\subsection{Monte-Carlo simulations
 \label{sec:coh3a}}
\begin{figure*}
\epsscale{1.}
\plotone{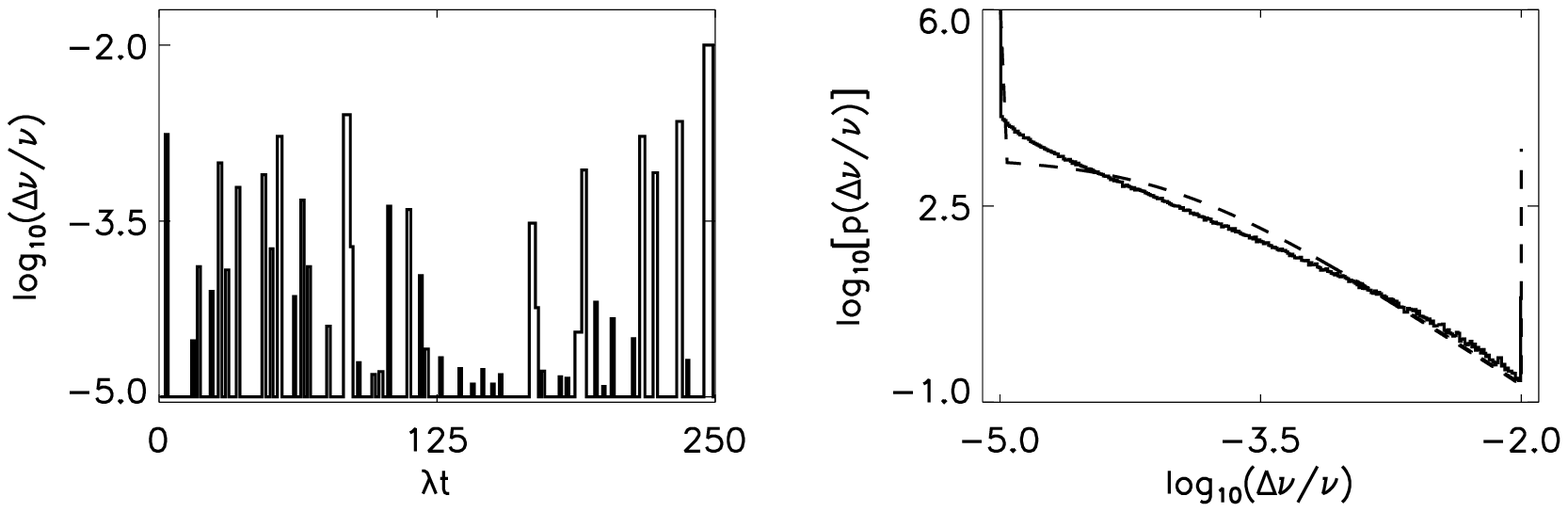}
\caption{
Sample output from a Monte-Carlo simulation of the coherent noise model
with $N=10^6$, $\epsilon=10^{-2}$, $f=10^{-3}$, $F_0 = 4\sigma$,
and $\Delta = 0.6 F_0$.
{\em Left panel.}
Time series of normalized glitch sizes, $\Delta\nu/\nu$, 
over the time interval $0\leq \lambda t \leq 255$.
{\em Right panel.}
Probability density function of normalized glitch sizes,
$p(\Delta\nu/\nu)$,
accumulated over the time interval $0 \leq \lambda t \leq 1\times 10^6$,
showing the simulation output (solid histogram)
and the analytic theory (dashed curve) overlaid.
The spikes at $\Delta\nu/\nu=10^{-5}$ (thermal creep only)
and $10^{-2}$ (all vortices unpin) are real features,
as explained in \S\ref{sec:coh2d}.
}
\label{fig:coh1}
\end{figure*}

Figure \ref{fig:coh1} (left panel) displays a time series 
of the glitches generated by the automaton 
over the interval $0\leq \lambda t \leq 255$
for 
$N=10^6$, $f=10^{-3}$, $\epsilon=10^{-2}$, $F_0=4\sigma$, and $\Delta = 0.6F_0$.
The glitches occur intermittently, with a Poissonian waiting-time distribution,
as arranged by construction through (\ref{eq:coh2}).
Of the 255 events appearing in the left panel of Figure\ \ref{fig:coh1},
most involve thermal unpinning only
($F_{\rm M}\leq F_0-\Delta$, $s=\epsilon f=10^{-5}$).
However, there are 45 events where the Magnus force exceeds
the pinning threshold at some of the occupied pinning sites
($F_{\rm M} > F_0-\Delta$),
including one where the Magnus force unpins every vortex
($F_{\rm M} \geq F_0+\Delta$, $s=\epsilon=10^{-2}$).

A frequency histogram of glitch sizes over a longer time interval
($0\leq \lambda t \leq 1\times 10^6$, corresponding to $10^6$ events) 
is constructed for the same parameters
and plotted on a log-log scale in Figure\ \ref{fig:coh1} 
(right panel, solid curve).
The analytic form of $h(s)$ derived from (\ref{eq:coh5})--(\ref{eq:coh7})
is plotted as a dashed curve over the simulation output.
The two curves agree at the upper end of the $s$ range,
but are significantly discrepant at the lower end,
because the time-averaged analytic theory in \S\ref{sec:coh2c}
does not capture the excess of small glitches arising from
temporal correlations (aftershocks; see \S\ref{sec:coh4}).
The discrepancies are smallest
for $\Delta \sim F_0$ and $F_0 \gtrsim \sigma$,
the regime of interest when fitting to pulsar data.

In the interval $-4.5 \lesssim \log_{10}\sigma \lesssim -2.5$,
$h(s)$ is a power law with exponent $-1.43\pm 0.01$
(weighted least squares fit).
In addition, $h(s)$ boasts two spikes
at $s=\epsilon f$ and $s=\epsilon$,
containing fractions 
$1 - e^{-(F_0-\Delta)/\sigma} \approx 0.80$
and
$e^{-(F_0+\Delta)/\sigma} \approx 1.7\times 10^{-3}$ 
of the total number of glitches respectively,
as discussed in \S\ref{sec:coh2d}.
The power law, which does not encompass the spikes at the upper and lower end of the glitch size distribution, is generic:
in the regime $F_{\rm p} \ll -\sigma \ln f$,
where forced unpinning dominates thermal creep,
we have $g(F_{\rm p}) \propto \exp(F_{\rm p})$,
$s(F_{\rm M}) \propto \exp(F_{\rm M})$
for $F_{\rm M} \gg F_0 - \Delta$,
and hence $h(s) \propto \exp[-2F_{\rm M}(s)] \propto s^{-2}$
asymptotically
\citep{sne97}.
Physically, the power law emerges as a historical effect,
because the system has memory.
For example, a middling value of $F_{\rm M}$ (say, $F_{\rm M} = F_0$)
may trigger a middle-sized glitch (say, $s\approx\epsilon/2$),
if $g(F_{\rm p})$ is fairly flat
(e.g.\ if every vortex unpinned during the previous iteration
of the automaton).
But the same middling value of $F_{\rm M}$ may trigger a tiny glitch 
instead (say, $\epsilon f \leq s \ll \epsilon/2$),
if $g(F_{\rm p})$ is grossly depleted in the range 
$F_0-\Delta \leq F_{\rm p} \lesssim F_0$
(e.g.\ following a chance sequence of iterations
with $F_{\rm M} \approx F_0$).
The latter outcome is more probable than the former
without preferring any particular glitch size,
so $h(s)$ scales as an inverse power of $s$ over a large portion 
of its domain.

\subsection{Shape of $h(s)$
 \label{sec:coh3b}}
\begin{figure*}
\epsscale{1.0}
\plotone{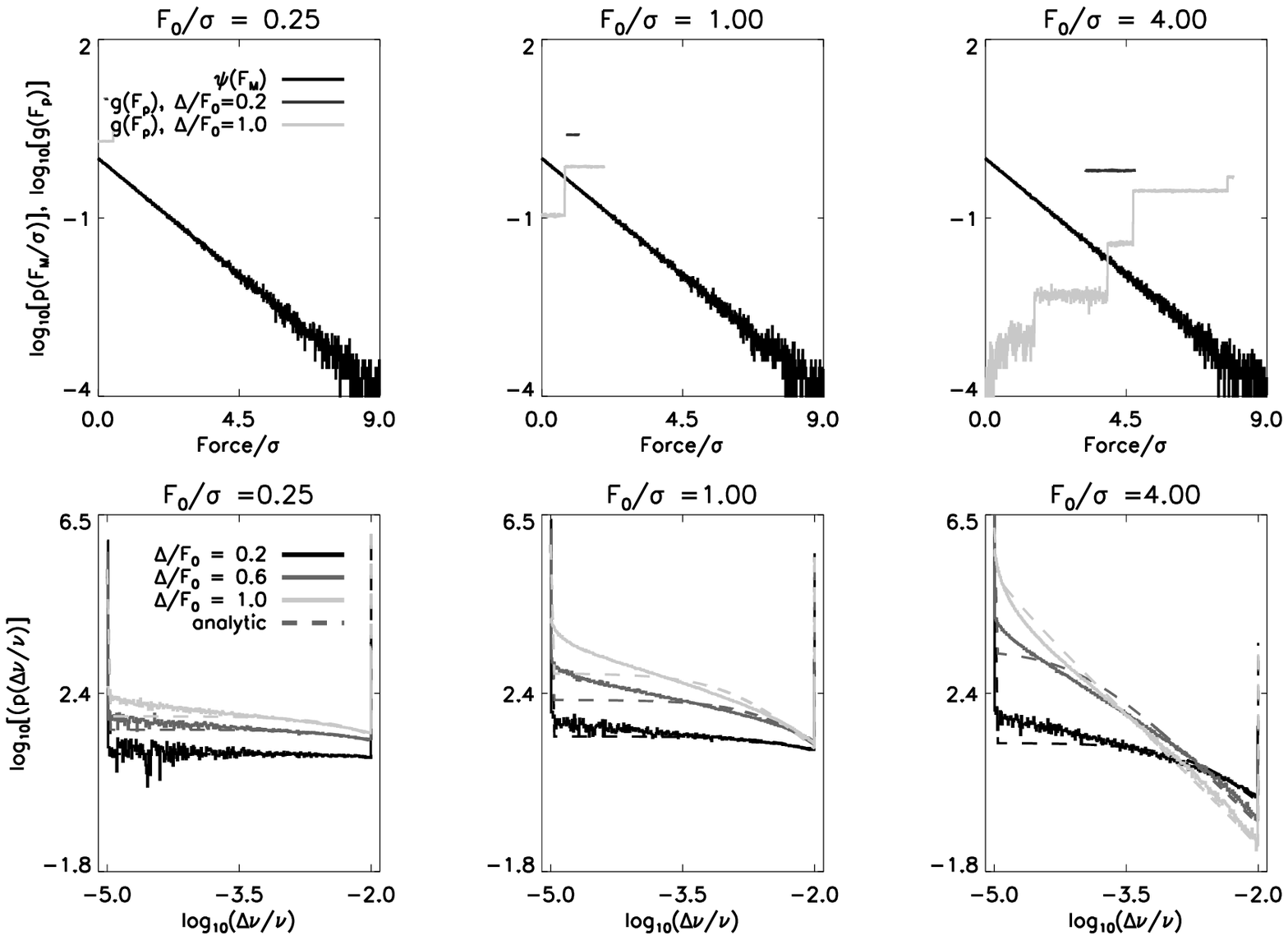}
\caption{
Theoretical size and pinning threshold distributions 
as functions of the pinning parameters $F_0$ and $\Delta$.
{\em Upper panels.}
Probability density functions of the Magnus stresses $\psi(F_{\rm M})$
(black histogram)
and pinning thresholds $g(F_{\rm p})$ (grey histogram),
for $\Delta/F_0 = 0.2$ (dark grey) and 1.0 (light grey),
and for
$F_0/\sigma= 0.25$ (left column), 1.0 (middle column), and 4.0 (right column).
{\em Lower panels.}
Probability density function of glitch sizes
for $\Delta/F_0 = 0.2$ (black), 0.6 (dark grey), and 1.0 (light grey),
and for
$F_0/\sigma= 0.25$ (left column), 1.0 (middle column), and 4.0 (right column).
The simulation output and analytic theory are graphed as
solid histograms and dashed curves respectively.
}
\label{fig:coh2}
\end{figure*}

The coherent noise model is completely specified by four parameters,
which together fix the shape of $h(s)$:
$F_0/\sigma$, $\Delta/\sigma$, $\epsilon$, and $f$.
The minimum and maximum glitch sizes produced by the model
are given by $\epsilon f$ and $\epsilon$ respectively.
The roles of $F_0/\sigma$ and $\Delta/\sigma$ are more subtle.
In combination, the latter two parameters determine the extent of the 
scale-invariant portion of $h(s)$,
as well as its log-log slope.
Figure \ref{fig:coh2} illustrates the various distributions 
that arise as we vary $\Delta/F_0$ in the range $0.2 \leq \Delta/F_0 \leq 1.0$,
given $F_0=0.25\sigma$ (left),
$F_0=1.0\sigma$ (middle),
and
$F_0=4.0\sigma$ (right).
There are two panels for each value of $F_0$.
The lower panel, which is drawn with a log-log scale,
displays $h(s)$ for
$\Delta = 0.2F_0$ (dark grey),
$\Delta = 0.6F_0$ (medium grey),
and
$\Delta = 1.0F_0$ (light grey),
showing the simulation output (solid histogram) and
analytic result (dashed curve) 
from Appendix \ref{sec:cohappa} superposed.
The upper panel, which is drawn with a log-linear scale,
displays $\psi(F_{\rm M})$ as a black histogram,
together with $g(F_{\rm p})$ for the minimum and maximum values of $\Delta/F_0$
(color coded as in the upper panel).
We run each simulation to get $10^6$ events.
The distributions in the upper panels are snapshots at the end
of the run, binned in units of $0.01\sigma$.
The step in $g(F_{\rm p})$ marks the last value of $F_{\rm M}$
sampled from $\psi(F_{\rm M})$ before the end of the run
(followed by random repinning).
The distributions in the lower panels are built up over the 
entire run,
binned in units of $0.01\,{\rm dex}$.

Certain trends in the shape of $h(s)$ are evident from Figure\ \ref{sec:coh2}.
(i) 
The distribution steepens as $F_0/\sigma$ increases,
while $\Delta/F_0$ is held constant, and vice versa.
(ii)
The probability density of the smallest glitches
excluding the left-hand spike, 
viz.\ $h(s\rightarrow \epsilon f^-)$,
increases as $F_0/\sigma$ increases,
while $\Delta/F_0$ is held constant, and vice versa.
In contrast,
the total probability in the left-hand spike increases
as $(F_0-\Delta)/\sigma$ increases.
(iii)
To maximize the scale-invariant portion of $h(s)$
and push the power-law exponent towards $2.0$,
we require $F_0 \gtrsim \sigma$ and $\Delta \sim F_0$.
(iv)
A gentle cusp appears in $h(s)$ at small $s$ 
for $\Delta \gtrsim 0.6\sigma$,
rendering the cumulative probability distribution
$\int_{\epsilon f}^s ds' \, h(s')$ more convex.

The above trends are related and easy to understand physically.
To get a steep power law, the memory effect described earlier
[i.e.\ gradual depletion of $g(F_{\rm p})$ over part of its domain]
must be free to take hold.
Yet it cannot do so properly if the threshold distribution is narrow
(which happens when $\Delta \ll F_0$),
or if large values of the Magnus force ($F_{\rm M} > F_0 + \Delta$)
occur frequently and reset $g(F_{\rm p})$ before it can be depleted
(which happens when $F_0 \ll \sigma$).
The trends are also easy to understand analytically.
For small $s$, 
where thermal creep dominates forced unpinning,
equations (\ref{eq:coh5})--(\ref{eq:coh7}) from the time-averaged theory
imply that $h(s)$ is flat,
whereas the full simulation (which preserves temporal correlations)
predicts $h'(s) < 0$ in the limit $s\rightarrow \epsilon f$,
reflecting the enhanced incidence of small aftershocks following a
large event (see \S\ref{sec:coh4}).
Elsewhere, in the regime $F_{\rm p} \ll -\sigma \ln f$,
where forced unpinning dominates thermal creep,
(\ref{eq:coh5})--(\ref{eq:coh7}) reduce to (\ref{eq:cohappa5}),
as shown in Appendix \ref{sec:cohappa}.
From (\ref{eq:cohappa5}), it is clear that the turnover
to a power law occurs for 
$s \gtrsim \epsilon (e^{2\Delta}-1)^{-1}$;
as a corollary, a power-law portion of $h(s)$ only develops
if we have $\Delta > 0.35\sigma$.
This is confirmed by Figure\ \ref{fig:coh2}.
For example,
you can see the turnover at $\log_{10} s \approx -2.6$ ($-4.1$)
in the dark (medium) grey curves in the right panel of Figure\ \ref{fig:coh2}
and at $\log_{10} s \approx -2.8$ in the light grey curve
in the middle panel.
The light grey curve in the right panel turns upwards
at $\log_{10} s \lesssim -2.8$ because condition (\ref{eq:cohappa4})
on the smallness of $f$,
which must be met to achieve a power law,
is violated for $f=10^{-3}$ and $F_0=\Delta = 4\sigma$.

\subsection{Extracting pinning parameters from observational data
 \label{sec:coh3c}}
We are now equipped to fit the coherent noise model to glitch data
from individual pulsars, in an effort to constrain the
fundamental parameters
$F_0/\sigma$, $\Delta/\sigma$, $\epsilon$, and $f$.
By way of preparation, we note three things.
First, $\lambda$ and hence $\sigma$ are directly measurable
from waiting-time data.
Second, the results in \S\ref{sec:coh3a} imply that any observed
size distribution is reproduced in shape by a 
relatively compact set of $F_0$ and $\Delta$ values,
which is encouraging.
Third, the sizes of the smallest and largest glitches observed,
denoted by $(\Delta\nu/\nu)_{\rm min}$
and $(\Delta\nu/\nu)_{\rm max}$ respectively,
constrain $\epsilon$ and $f$ according to
$(\Delta\nu/\nu)_{\rm max} \leq \epsilon \leq 1$
and
$N^{-1} \leq f \leq (\Delta\nu/\nu)_{\rm min} / (\Delta\nu/\nu)_{\rm max}$.
Of course, if $\epsilon$ is much larger than $(\Delta\nu/\nu)_{\rm max}$,
we expect to see glitches larger than 
$(\Delta\nu/\nu)_{\rm max}$ if we observe for longer.
Luckily, this missing information
is not a serious problem when fitting the model to the data,
because
$\int_{(\Delta\nu/\nu)_{\rm max}}^\epsilon ds' \, h(s')$
is usually small.
On the other hand, it is also possible that $f$ is much smaller than
$(\Delta\nu/\nu)_{\rm min} / (\Delta\nu/\nu)_{\rm max}$,
yet we fail to see glitches smaller than $(\Delta\nu/\nu)_{\rm min}$
because the resolution of pulsar timing experiments is imperfect.
This problem is more serious,
because
$\int_{\epsilon f}^{(\Delta\nu/\nu)_{\rm min}} ds' \, h(s')$ 
is usually large;
much, perhaps most, of the distribution may be invisible.
To handle the problem optimally,
one should quantify the resolution experimentally for each individual pulsar
\citep{jan06},
derive an observational window function $w(s)$,
and fit the data to $h(s) w(s)$.
Calibrated window functions are not published for most pulsars, 
so implementing the foregoing procedure lies outside the scope of this paper.
Instead, we proceed conservatively
by tailoring our fits to the observed bounds according to
$\epsilon=(\Delta\nu/\nu)_{\rm max}$
and
$f = (\Delta\nu/\nu)_{\rm min} / (\Delta\nu/\nu)_{\rm max}$.
The results are interpreted critically in \S\ref{sec:coh4}
in light of the uncertainties flagged above.

\begin{figure*}
\epsscale{2.25}
\plottwo{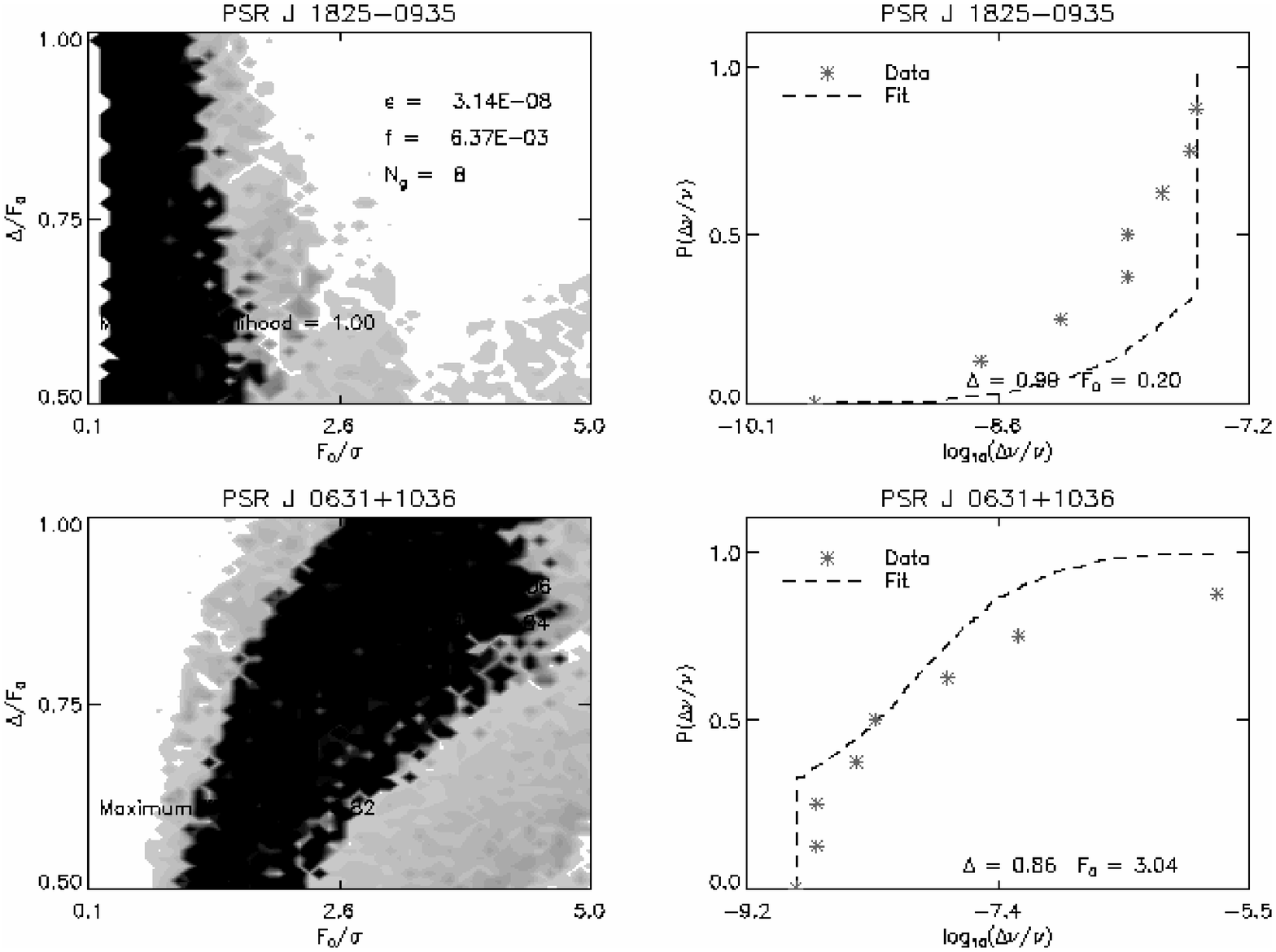}{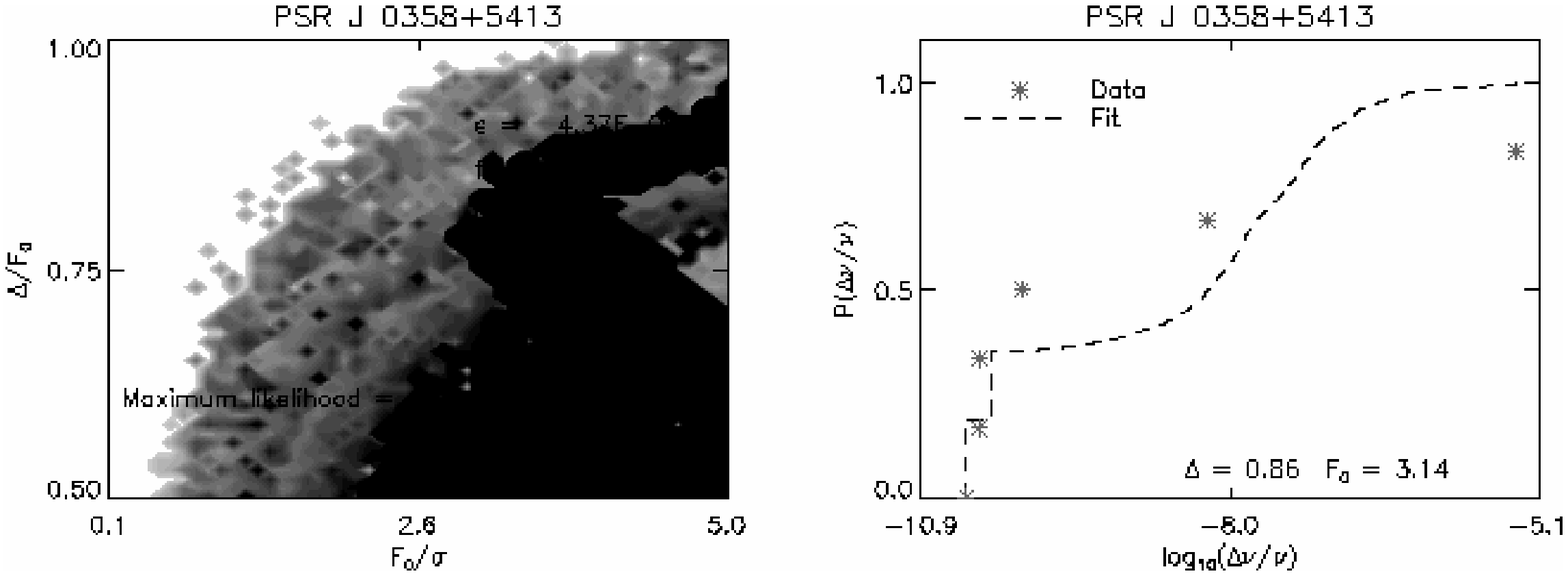}
\caption{
Fits of the coherent noise model to pulsar glitch data for the seven pulsars that 
have glitched at least six times, and whose waiting-time distributions do not show 
a discernible periodic component.  
{\em Left column.} 
Greyscale plots of the relative likelihood function for combinations of 
$F_0/\sigma$ and $\Delta/F_{0}$ 
in the ranges $0.1$--$5.0$ and $0.5$--$1.0$ respectively.
The grey scale runs from least (white) to greatest (black) likelihood.  
The values of $\epsilon$ and $f$ used to generate the fits
are printed on each panel, 
together with the maximum relative likelihood for the best fit.  
{\em Right column.} 
Cumulative probability function for the best fit (dashed curve),
plotted over the measured cumulative probability function (asterisks).
}
\label{fig:coh4}
\end{figure*}

In Figure\ \ref{fig:coh4},
we fit the model to measurements of $h(s)$ in the seven pulsars
which have glitched more than five times
without a discernible periodic component
\citep{mel08},
i.e.\ with a purely Poissonian waiting-time distribution,
as assumed by the model through (\ref{eq:coh2}).
The fits are done using a maximum likelihood approach,
described in Appendix \ref{sec:cohappb},
and the fitted parameters are recorded in Table \ref{tab:coh1}.
There are two panels attached to each object in Figure\ \ref{fig:coh4}.
In the left panel, the likelihood function ${\cal L}$ is presented as a 
greyscale plot in the $F_0/\sigma$-$\Delta/F_0$ plane,
such that larger ${\cal L}$ values are shaded darker than smaller ${\cal L}$ values.
In the right panel, the continuous cumulative distribution
$\int_{\epsilon f}^s ds' \, h(s')$ corresponding to the 
maximum likelihood --- the best fit ---
is plotted as a dashed curve, together with the observational data (asterisks).  It should be noted that the delta-function peaks that appear in the differential probability distributions at the upper and lower bounds on the glitch size appear as steps in the cumulative distributions.  
The objects in Figure\ \ref{fig:coh4}
are ranked in decreasing order of the number of glitches observed, $N_{\rm g}$.

\begin{figure*}
\epsscale{2.25}
\plottwo{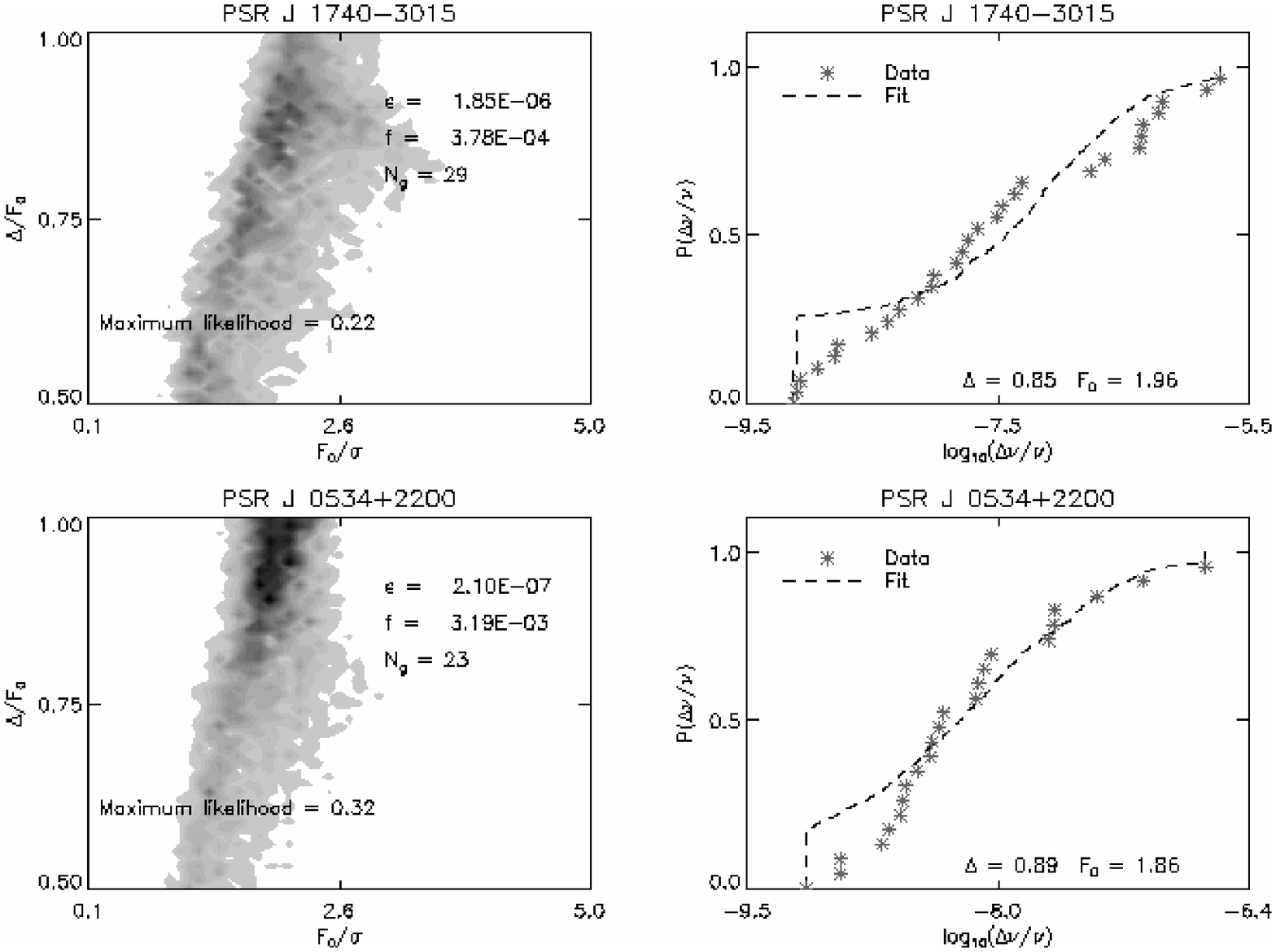} {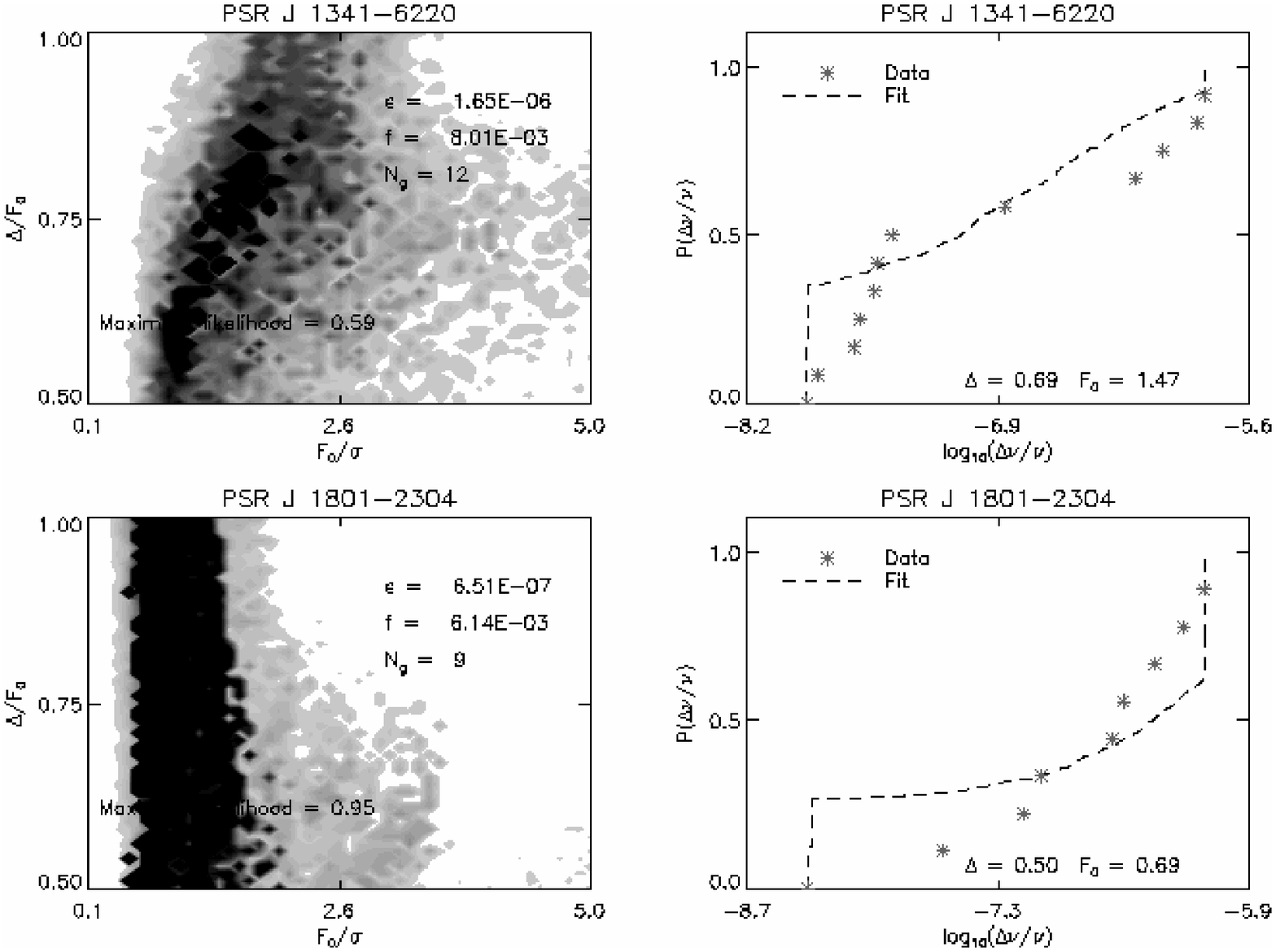}
\end{figure*}

\begin{table}
\begin{center}
\begin{tabular}{crrrlll}
\hline \hline
 PSR J & $N_{\rm g}$ & $\log_{10}\epsilon$ & $\log_{10} f$ & 
   $F_0/\sigma$ & $\Delta/F_0$ & ${\cal L}$ \\ \hline
 1740$-$3015 & $29$ & $-5.7$ & $-3.4$ & 2.0 & 0.87 & 0.22 \\
 0534$+$2200 & $23$ & $-6.7$ & $-2.5$ & 1.9 & 0.89 & 0.32 \\
 1341$-$6220 & $12$ & $-5.8$ & $-2.1$ & 0.49 & 0.12 & 1.0 \\ 
 1801$-$2304 & $9$  & $-6.2$ & $-2.2$ & 0.30 & 0.23 & 1.0 \\
 1825$-$0935 & $8$  & $-7.5$ & $-2.2$ & 0.20 & 0.75 & 1.0 \\
 0631$+$1036 & $8$  & $-5.8$ & $-2.9$ & 3.0 & 0.86 & 0.82 \\ 
 0358$+$5413 & $6$  & $-5.4$ & $-5.2$ & 3.1 & 0.86 & 1.0 \\ \hline
 0537$-$6910 & $23$ & $-6.2$ & $-1.6$ & 0.84 & 0.23 & 0.03 \\
 0835$-$4510 & $17$ & $-5.5$ & $-2.7$ & 0.59 & 0.95 & 0.11 \\ 
\hline \hline
\end{tabular}
\end{center}
\caption{
Pinning parameters extracted from the maximum likelihood fits 
in Figures \ref{fig:coh4} and \ref{fig:coh6}.
}
\label{tab:coh1}
\end{table}

PSR J0534$+$2200 and PSR J1740$-$3015 are fitted well
by the coherent noise model.
For both objects, the best fit is achieved for
$F_0 \approx 2\sigma$ and $\Delta\approx 0.9F_0$.
The other objects are fitted reasonably well too,
allowing for the simplicity of the model.
For example, the fit for PSR J1825$-$0935 
(fifth row, left panel, Figure\ \ref{fig:coh4})
looks mediocre to the eye.
But the likelihood function (right panel) is quite bumpy,
with several peaks of almost equal heights,
some of which
(e.g.\ at $\Delta\approx 0.9F_0$)
produce a fit which looks better to the eye,
at the expense of a marginal drop in likelihood.
For all seven objects,
the best fits are achieved for
$\Delta \gtrsim 0.5 F_0$ and $0.8\lesssim F_0/\sigma \lesssim 3$.
This finding is in accord with the theoretical analysis
in \S\ref{sec:coh3a}:
one requires
$F_0 \gtrsim \sigma$ and $\Delta\sim F_0$ 
in order to get a power-law size distribution
$h(s) \propto s^a$ that is steep enough
($-2 \lesssim a \lesssim -1$) to match the observations
\citep{jan06,mel08}.
In addition, one requires 
$F_0 + \Delta \gtrsim -\sigma \ln f
 \approx (5$$-$$7)\sigma$
in order to get a cumulative distribution 
$\int_{\epsilon f}^s ds' \, h(s')$ that is concave up.

Some of the tightest constraints come from the spike 
in $h(s)$ at $s=\epsilon f$.
Four objects
(PSR J0534$+$2200, PSR J1740$-$3015, PSR J1801$-$2304,
and PSR J1825$-$0935)
exhibit no evidence of a spike at $s\approx (\Delta\nu/\nu)_{\rm min}$
in the data,
implying either that these objects have $\Delta\approx F_0$,
or that $(\Delta\nu/\nu)_{\rm min}$ comfortably exceeds $\epsilon f$
because the experimental resolution prevents smaller glitches
from being seen.
On the other hand, the remaining three objects
(PSR J0358$+$5413, PSR J0631$+$1036, and PSR J1341$-$6220)
do exhibit some evidence of a spike at $s\approx (\Delta\nu/\nu)_{\rm min}$,
implying $\Delta \lesssim 0.6 F_0$ and that the smallest glitches
are resolved.
By contrast, the spike at $s=\epsilon$ does not constrain the fits tightly
because it does not contain much integrated probability.

The objects in Figure\ \ref{fig:coh4} and Table \ref{tab:coh1}
are ranked in order of decreasing $N_{\rm g}$ to make the 
obvious but important point that the localization of the peak
in the likelihood function, and hence the significance of the fit,
improve markedly as $N_{\rm g}$ increases.
Encouragingly, the model also fits the data better as $N_{\rm g}$ increases.
The next task is to test the model more stringently by finding more glitches,
both by reanalyzing timing data from the seven pulsars
in Figure\ \ref{fig:coh4} to search for small, overlooked events,
and by undertaking high-duty-cycle timing campaigns in the future.

\subsection{Quasiperiodic glitchers
 \label{sec:coh3d}}
The timing histories of PSR J0537$-$6910 ($N_{\rm g}=23$) and  PSR J0835$-$4510 (Vela; $N_{\rm g}=17$) harbor a periodic glitching component,  which coexists with the scale-invariant component discussed so far \citep{lyn96,mar04,mid06,mel08}. The periodic component comprises $\sim 25$ per cent  of events and is characterized by narrowly peaked  size and waiting-time distributions. When we attempt to fit the theoretical $h(s)$ given by (\ref{eq:cohappa3}) to the data from these two objects, following the same procedure as in Figure\ \ref{fig:coh4}  and Appendix \ref{sec:cohappb}, the agreement is no better or worse than for a pure Poissonian glitcher. Taken at face value, this is surprising; $\psi(F_{\rm M})$, a basic input into the model, is wrongly specified by (\ref{eq:coh2}) for quasiperiodic glitchers, because it does not contain a narrowly peaked component. However, we show below that this omission does not show up noticeably in the observed $h(s)$ for small $N_{\rm g}$ 
(e.g.\ $N_{\rm g}\leq 23$).

\begin{figure*}[ht!]
\epsscale{1.0}
\plotone{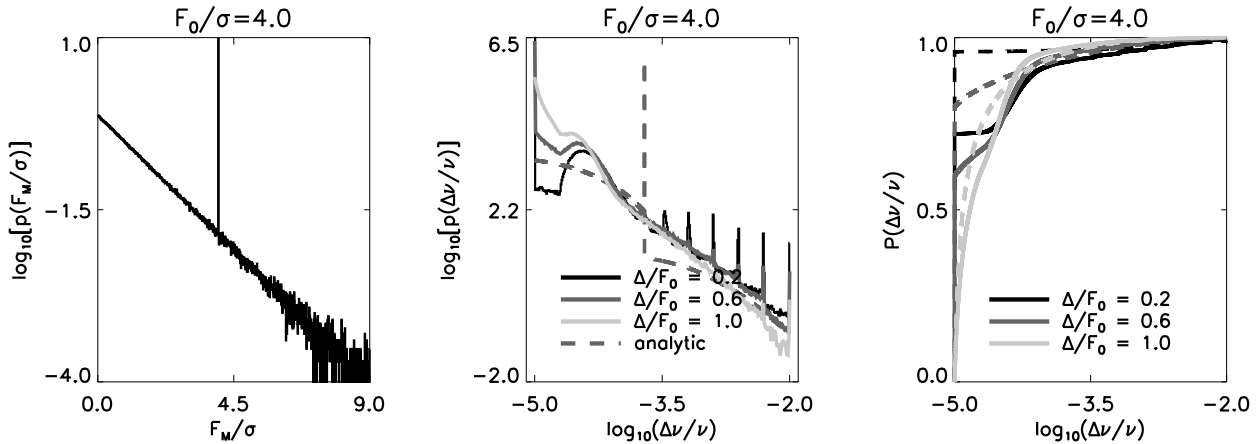} 
\caption{
Output of the coherent noise model including a periodic component 
weighted at $25\%$,
with
$N=10^6$, $\epsilon=10^{-2}$, $f=10^{-3}$,
$F_0 = 4.0\sigma$, $F_{\rm q} = 4.0\sigma$, 
and
$\Delta/F_0 = 0.2$ (light grey), $0.6$ (dark grey), or $1.0$ (black).  
{\em Left panel.}
Magnus force probability density function, $\psi(F_{\rm M})$.
The spike at $F_{\rm M}=F_{\rm q}$ contains $\approx 25\%$ of the events.
{\em Middle panel.}
Glitch size probability density function $h(s)$,
featuring the output of the Monte-Carlo simulations (solid histogram)
and the analytic theory for $\Delta=0.6 F_0$ (dashed curve).
Note the row of spikes generated by consecutive periodic events,
whose $s$-positions and heights form geometric sequences with
common ratios $q$ and $C_{\rm q}$ respectively.
Note also the hump at low $s$,
which makes the cumulative distribution concave at low $s$.
{\em Right panel.}
Cumulative size distribution $\int_{\epsilon f}^s ds' \, h(s')$ 
for the models in the middle panel (matching color code).
The solid (dashed) curves refer to models that include (exclude)
the periodic component.
}
\label{fig:coh5}
\end{figure*}

We recalculate the coherent noise model for
$\psi(F_{\rm M}) = 
 (1-C_{\rm q}) \sigma^{-1} \exp(-F_{\rm M}/\sigma)
 + C_{\rm q} \sigma^{-1} \delta(F_{\rm M}/\sigma - F_{\rm q}/\sigma)$,
where
$F_{\rm q}$
is the Magnus force built up during one period $t_{\rm q}$,
and $C_{\rm q}$ defines the periodic fraction;
see \S5.2 of \citet{mel08}.
Figure \ref{fig:coh5} illustrates how the coherent noise process
operates in the interesting case where $F_{\rm q}$ lies within the domain 
$|F_{\rm p} - F_0| \leq \Delta$,
where $g(F_{\rm p})$ is nonzero.
\footnote{
For $F_{\rm q} \leq F_0 - \Delta$, the results in \S\ref{sec:coh3a}
and \S\ref{sec:coh3b} carry over without change, 
because the periodic component does not unpin anything,
and the number of glitches with $s > \epsilon f$
decreases by a factor $C_{\rm q}$.
For $F_{\rm q} \geq F_0 + \Delta$, 
the periodic component unpins every vortex, 
and the number of glitches with $s=\epsilon$ increases 
by a factor $C_{\rm q}$,
while the rest of $h(s)$ remains unchanged.
}
The model parameters are 
$N=10^6$, $\epsilon=10^{-2}$, $f=10^{-3}$,
$F_0 = 4.0\sigma$, $F_{\rm q} = 4.0\sigma$, and $C_{\rm q}=0.25$.
The left panel of Figure\ \ref{fig:coh5}
displays $\psi(F_{\rm M})$ on a log-linear scale;
the spike housing the periodic component is clearly visible.
The middle panel displays $h(s)$ from the simulation
on a log-log scale for
$\Delta=0.2F_0$ (dark grey histogram),
$\Delta=0.6F_0$ (medium grey histogram),
and
$\Delta=1.0F_0$ (light grey histogram),
with the analytic prediction for $\Delta=0.6 F_0$ overlaid as a dashed curve.
The right panel displays $\int_{\epsilon f}^s ds' \, h(s')$ 
on a linear-log scale for the same three values of $\Delta$
(color coded as in the middle panel),
showing the simulation output including (solid curves) 
and excluding (dashed curves) the periodic component.
One immediately sees that, 
for $\Delta = 0.6\sigma$ and $1.0\sigma$,
where $F_{\rm q}$ lies within the domain of $g(F_{\rm p})$,
there is a turnover in $h(s)$ at $s(F_{\rm q})$.
Additionally,
if the probability of getting $F_{\rm M} = F_0 - \Delta$
greatly exceeds that of getting $F_{\rm M}=F_0 + \Delta$,
such that thermal creep unpins more sites on average
than events with $F_{\rm M}\geq F_0 + \Delta$,
then condition (\ref{eq:cohappa4}) is violated,
the power-law form of $h(s)$ breaks down,
and an excess (i.e.\ hump) of glitches emerges at small $s$.

The main new feature in $h(s)$ is a sequence of spikes at 
$s=q\epsilon$, $q^2\epsilon$, $q^3\epsilon$, $\dots$,
with $q = (F_q - F_0 + \Delta) / (2\Delta)$,
whose heights are in the ratio $1:C_{\rm q} : C_{\rm q}^2:\dots$.
The spikes correspond to one, two, three, $\dots$ consecutive draws
from the periodic peak at $F_{\rm M} = F_{\rm q}$ in $\psi(F_{\rm M})$,
with each draw unpinning a fraction $\approx q$ of the vortices,
of which a fraction $q$ subsequently repin with $F_{\rm p} \leq F_{\rm q}$.
The spikes in the simulation output do indeed occur at the predicted positions,
although, apart from the first few, they are hard to see,
swamped by the flood of Poissonian low-$s$ events.
On the other hand, the analytic, mean-field theory only predicts one spike,
at $s=q\epsilon$;
as a time-averaged theory (see \S\ref{sec:coh2c}),
it does not recognize correlated event sequences
(e.g.\ consecutive $F_{\rm M}= F_{\rm q}$ draws) as special.
For example, the analytic theory does not distinguish between
two Poisson events followed or separated by a periodic event,
yet the relative frequency of these sequences determines the shape
of the broad turnover in $g(F_{\rm p})$ at $F_{\rm p}=F_{\rm q}$
(where the analytic model predicts a simple step).

\begin{figure*}[ht!]
\epsscale{1.0}
\plotone{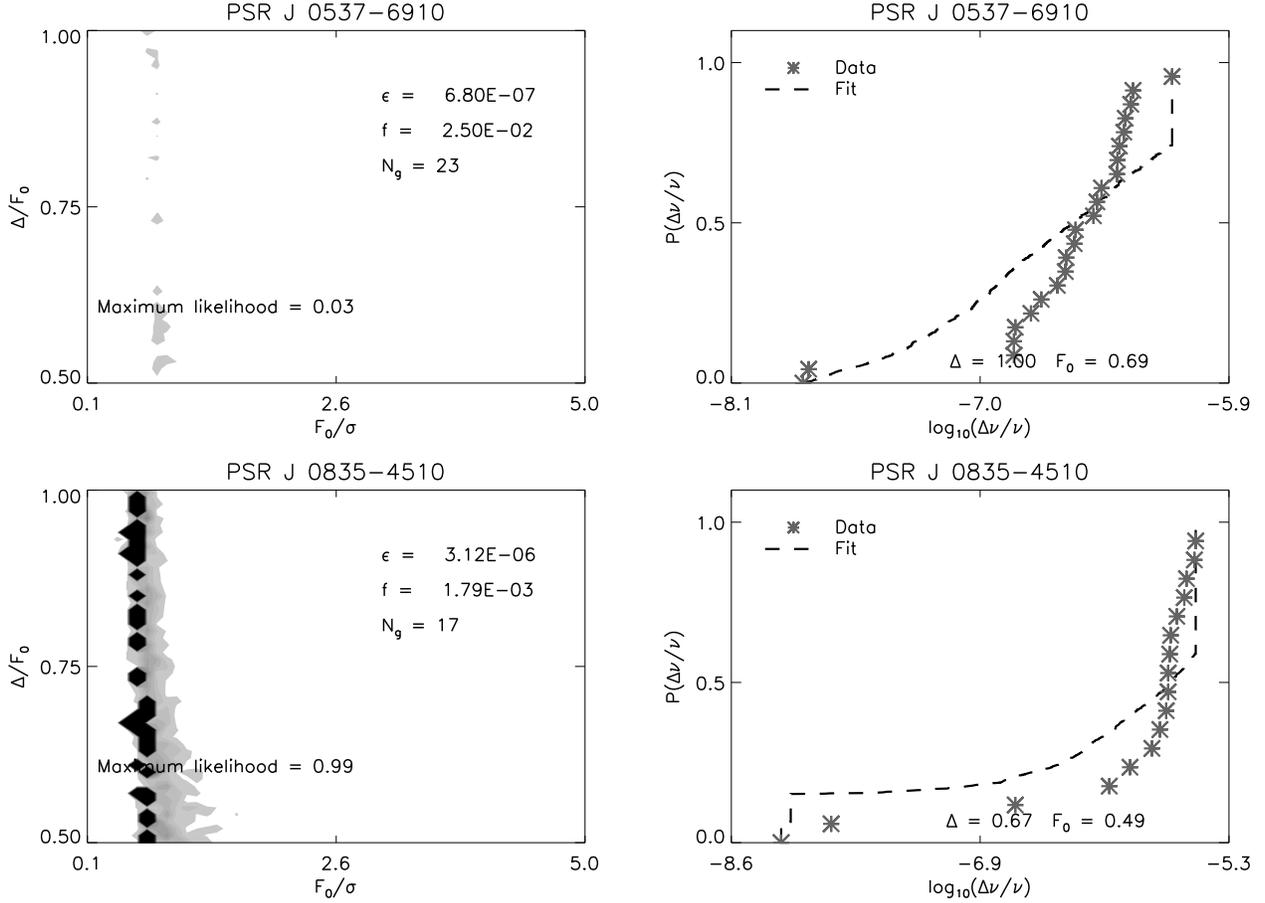}
\caption{
Fits of the coherent noise model to glitch data for PSR J0537$-$6910
and PSR J0835$-$4510, whose waiting-time distributions feature
a periodic component.
{\em Left column.} 
Greyscale plots of the relative likelihood function for combinations of 
$F_0/\sigma$ and $\Delta/F_{0}$ 
in the ranges $0.1$--$5.0$ and $0.5$--$1.0$ respectively,
in the context of a purely Poissonian model.
The grey scale runs from least (white) to greatest (black) likelihood.  
The values of $\epsilon$ and $f$ used to generate the fits
are printed on each panel, 
together with the maximum relative likelihood for the best fit.  
{\em Right column.} 
Cumulative probability function for the best fit (dashed curve),
plotted over the measured cumulative probability function (asterisks).
}
\label{fig:coh6}
\end{figure*}

The spikes are a key prediction of the coherent noise model
for quasiperiodic glitchers. Do they show up in the data
(as steps in the cumulative size distribution)?
No; nor should we expect them to,
when so few glitches have been detected,
and each spike contains modest probability.
To date, the objects 
PSR J0537$-$6910 and PSR J0835$-$4510
have been seen to sample the underlying event distribution
$N_{\rm g}=23$ and $N_{\rm g}=17$ times respectively.
Monte-Carlo realizations of the coherent noise model
with $C_{\rm q}=0$ and $C_{\rm q}=0.25$ are statistically indistinguishable
for $N_{\rm g}=23$ (and hence $N_{\rm g}=17$).
Consequently, we experience no loss of generality if we fit a
purely Poissonian model to the data
from PSR J0537$-$6910 and PSR J0835$-$4510,
following the same procedure as in Figure\ \ref{fig:coh4}
and Appendix \ref{sec:cohappb}.
The results are graphed in Figure\ \ref{fig:coh6}
(copying the format of Figure\ \ref{fig:coh4}),
and the best fit parameters are recorded in the lower part 
of Table \ref{tab:coh1}.
A respectable fit is achieved for PSR J0835$-$4510,
with $F_0\gtrsim \sigma$ and $\Delta \approx F_0$ as usual,
but not for PSR J0537$-$6910, 
whose measured size distribution is skewed by two isolated glitches
with $s\approx 10^{-7.8}$ 
and a large group of glitches centered at $s\approx 10^{-6.5}$
(implying, curiously, the existence of a larger periodic component than
the observed waiting-time distribution can accommodate).
The results raise the possibility that some small glitches were missed
when reducing timing data from PSR J0537$-$6910.
As matters stand, the data are equally consistent with the presence or absence
of periodic spikes.
We need to detect more glitches with better resolution
to settle the issue conclusively.

\section{Discussion
 \label{sec:coh4}}
In this paper, we propose an idealized model of pulsar glitches,
in which the collective unpinning of superfluid vortices
is treated as a coherent noise process \citep{sne97}.
The model accounts for the scale invariance of glitch sizes over the range of possible glitch sizes,
through the interplay between thermal and forced unpinning,
with the two mechanisms balancing each other in a time-averaged sense.
It reproduces approximately the measured size distributions 
of the nine most active glitchers
for a range of physically sensible pinning threshold distributions,
despite grossly oversimplifying the microphysics.
More importantly, it demonstrates a fundamental and counterintuitive
point of principle,
which holds irrespective of the detailed microphysics:
scale-invariant glitches can arise
from {\em homogeneous} collective unpinning.
Coherent noise is therefore a viable alternative to the large family
of inhomogeneous mechanisms proposed in the literature
(e.g.\ self-organized criticality),
which rely upon nearest-neighbor avalanches and the existence
of large-scale capacitive regions.

Such an alternative is welcome.
It fits with comtemporary notions of nuclear pinning and thermal creep,
where it is thought that vortices hop between adjacent (or nearly adjacent) lattice pinnning sites,
which lie $\sim 30\,{\rm fm}$ apart and are therefore distributed homogeneously
on relevant macroscopic scales, e.g.\ the period of the Abrikosov lattice
\citep{alp84,jon91,lin93,jah06}.
It also helps to explain the puzzle, posed in \S\ref{sec:coh1},
of why glitch sizes vary so much in an individual pulsar,
even though the global Magnus stress builds up to a similar level 
between glitches (dictated by the Poissonian waiting-time distribution)
and is felt simultaneously by every pinned vortex.

\subsection{$F_0$ and $\Delta$
 \label{sec:coh4a}}
Disaggregated glitch statistics for individual pulsars became available
in meaningful volume only recently, 
yet already they tell a clearer story than aggregate statistics
\citep{mel08}.
(i) When disaggregated, the waiting-time distribution is Poissonian, 
except in two pulsars which have a small ($\sim 25\%$) periodic component.
(ii) When disaggregated, the size distribution is consistent with a power law
$h(s)\propto s^a$,
but the power-law exponent $a$ is not universal;
the aggregate size distribution is inconsistent with a sum of 
identically sloped power laws at the 99\% level of confidence
[see \S4.4 of \citet{mel08}].
(iii) The observed cumulative size distribution 
of the Poissonian glitchers
is concave in two objects, convex in two others,
and contains a point of inflexion in the remaining three.

The glitch model in this paper is motivated by (i).
Its predictions are consistent with (ii) and (iii).
We find the following properties.
First,
the theoretical $h(s)$ is scale invariant over several decades in each object,
with $-2\leq a \leq 0$.
Second,
the observed absence of a step at $s=\epsilon f$ in most objects implies
a broad pinning threshold distribution,
with $\Delta \approx F_0$.
Physically, this conclusion is natural:
nuclear structure calculations in the local density and
Hartree-Fock-Bogoliubov approximations independently indicate
that the pinning energies in a neutron star covers a broad range
($1\,{\rm MeV} \lesssim E_{\rm p} \lesssim 4\,{\rm MeV}$),
because the superfluid density $\rho_{\rm s}$ changes with depth
\citep{lin91,jon97,jon98,don03,don04,don06,avo07},
translating into a broad $\phi(F_{\rm p})$ when we fit our
homogeneous model to data.
\footnote{
When vortices repin after a glitch, $g(F_{\rm p})$ is broad.  During thermal creep, which involves continuous angular momentum transfer from the superfluid to the crust, the vortices gradually bend in a piecewise fashion, via a sequence of small rearrangements which bring them into line with local microcrystalline boundaries. Consequently, $\langle F_{\rm p} \rangle$ increases and $g(F_{\rm p})$ narrows, until the next glitch intervenes. For a fuller exposition of this idea, please consult \citet{jon91}.
}
Third, most objects have $-2\lesssim a \lesssim -1$, i.e.\ $h(s)$ is not flat, which requires $F_0 \gtrsim \sigma$. Fourth, objects with a concave segment in $h(s)$ fall in the regime  where the rates of thermal and forced unpinning are comparable [i.e.\ condition (\ref{eq:cohappa4}) is violated], producing a hump in $h(s)$ at small $s$.

The coherent noise model also works fairly well for the quasiperiodic glitchers PSR J0537$-$6910 and PSR J0835$-$4510 and leads to similar conclusions,  i.e.\ $F_0\gtrsim \sigma$ and $\Delta\approx F_0$. In principle, if $\psi(F_{\rm M})$ contains a periodic component,  a sequence of equally spaced (in $\log_{10}s$) spikes appears in $h(s)$ (see \S\ref{sec:coh3d}). The available data are consistent with the spikes,  as Figure\ \ref{fig:coh6} shows, but one cannot hope to actually trace out the associated steps in the cumulative size distribution unless $N_{\rm g} \gtrsim 10^6$. The periodic component impacts more on the waiting-time distribution, which flattens at small $\Delta t$,  than on $h(s)$, which rises at small $s$, swamping the spikes. Crucially, quasiperiodic glitching does not discriminate between homogenoeus (e.g.\ coherent noise) and inhomogeneous (e.g.\ avalanche) collective unpinning; it is equally at home in both families of models. For example, in self-organized critical systems (e.g.\ sand piles), quasiperiodicity is caused by system-spanning avalanches triggered by a fast external driver \citep{jen98,sor04,mel08}. 

In order to exploit the results presented in Table \ref{tab:coh1} to infer an actual value of $F_{0}$, we must define $\sigma$ in terms of physical pulsar parameters.  We present two possible definitions of $\sigma$.  Firstly, we assume that the Magnus stress that accumulates between glitches is proportional to the time between glitches, such that 
\begin{equation}
 \sigma_1 = 2\pi \dot{\nu} R \rho_{\rm s} \kappa / \lambda~.
 \label{eq:sigma_delta_t}
\end{equation}
where $\lambda$ is the mean glitching rate, $\dot{\nu}$ is the rate of angular deceleration, $R$ is the characteristic radius at which vortex pinning occurs (or, more correctly, the characteristic radius where pinning occurs most strongly), and $\rho_{\rm s}$ is the superfluid mass density at that radius \citep{don06,avo07}.  The $F_{\rm{M}}-\Delta t$ correlation underlying Equation (\ref{eq:sigma_delta_t}) has been proposed often in the literature in the context of models without thermal creep \citep{and75,mor96,lyn00,wan00} and finds support in the reservoir effect observed in PSR J0537$-$6910 \citep{mid06}.  This is an opportune time to test its consequences. 

\begin{figure}[ht!]
\epsscale{1.0}
\begin{center}
\plotone{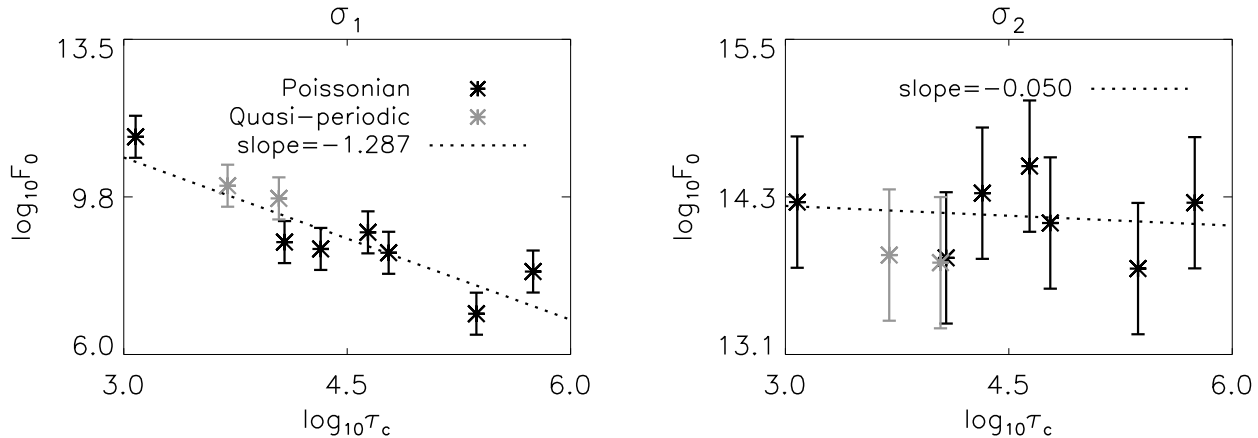}
\end{center}
\caption{
Best fit mean pinning threshold $F_{0}$ (units: ${\rm N\,m^{-1}}$) versus characteristic spin-down age $\tau_{\rm{c}}$ (units: kyr) for all nine pulsars that have glitched at least six times for two different definitions of $\sigma$:  $\sigma_1$ from Equation (\ref{eq:sigma_delta_t}) (left) and $\sigma_2$ from Equation (\ref{eq:sig2})(right).   The data are drawn from Table \ref{tab:coh1}. Quasiperiodic glitchers are plotted in grey.}
\label{fig:coh8}
\end{figure}

A key corollary of \ref{eq:sigma_delta_t} is that the average pinning strength $F_0 \sim \sigma$ varies greatly  between neutron stars. In the left panel of Figure\ \ref{fig:coh8}, we plot $F_0$ (in ${\rm N\,m^{-1}}$) versus characteristic spin-down age $\tau_{\rm c} = -\nu/(2\dot{\nu})$ (in kyr) for the nine pulsars in Table \ref{tab:coh1} assuming $\sigma = \sigma_1$. We estimate the error bars conservatively to be $\pm 0.5$ dex  from Figures \ref{fig:coh4} and \ref{fig:coh6} by inspection.  Remarkably, there is an inverse correlation over three decades in $\tau_{\rm c}$ and two decades in $F_0$, with $F_0 \propto \tau_{\rm c}^{-1.28 \pm 0.31}$ approximately. Taken at face value, the correlation is surprising physically: $F_0$ is determined by the structure of the nuclear lattice, averaged over position within the stellar crust in our model. It is therefore primarily a function of the mean crustal density, which varies slightly across the neutron star population. One might argue that Figure\ \ref{fig:coh8} indicates a temperature effect: as the star cools, thermal creep occurs more slowly. However, the rate of thermal creep is parametrized by $f$, not $F_0$. One might argue instead that $F_0$ depends sensitively on  $\rho_{\rm s}$ and hence $T$. However, this effect seems to work in the wrong direction; \citet{don06} found that $E_p$ increases as $T$ decreases  and $\rho_{\rm s}$ increases. We therefore speculate that the nuclear lattice ``anneals'' as the star ages, with weak seismic and/or thermal fluctuations erasing defects as time passes.  This process of annealing would have a less dramatic effect on the mean pinning energy if pinning is mainly due to sites in the nuclear lattice.  It is well known that the concentration of monovacancies affects the strength of pinning severely \citep{jon91,jon97,jon98}. But it is an open question if a realistic annealing process exists to lower the monovacancy concentration with time. 


An alternative definition of $\sigma$ invokes the process of thermal creep to define the mean differential lag between the interior superfluid and stellar crust \citep{alp84,lin91,jah06}, viz.
\begin{equation}
 \sigma_2 = \delta \omega R \rho_{\rm s} \kappa,
 \label{eq:sig2}
\end{equation}
with
\begin{equation}
 \delta \omega = \omega_{\rm cr} \frac{k_{\rm B}T}{E_{\rm p}}\ln\left(4\tau_{\rm C} v_0/R\right)~,
 \label{eq:domega}
\end{equation}
where $\omega_{\rm cr}\approx 0.1\,\rm{rad\,s}^{-1}$ is the critical lag above which vortex pinning can no longer be sustained, $k_{\rm B} T/E_{\rm p}\approx 30$ is the ratio of the typical thermal energy and pinning energies, and $v_0\approx 10^7\,\rm{ms}^{-1}$ is a trial microscopic vortex creep speed.  [Note that no self-consistent theory exists at present for the stochastic fluctuations of $\delta \omega$ from glitch to glitch about its mean (\ref{eq:domega}).]  When the fitted values of $F_0$ are recalculated using $\sigma = \sigma_2$ and plotted against $\tau_{\rm{c}}$ (right panel of Figure \ref{fig:coh8}), the inverse correlation observed in the left panel of Figure \ref{fig:coh8} is no longer present at a significant level.  In this case $F_0 \propto \tau_{\rm c}^{-0.05 \pm 0.14}$, suggesting that the mean vortex pinning strength does not change from pulsar to pulsar and therefore seems to be independent of pulsar temperature.  It should be noted, however, that both $\sigma_1$ and $\sigma_2$ are interpreted assuming that the characteristic radius at which pinning occurs does not depend on characteristic age (and hence temperature) and therefore it does not vary from pulsar to pulsar.

\subsection{$\epsilon$ and $f$
 \label{sec:coh4b}}
\begin{figure*}[ht!]
\epsscale{1.0}
\plotone{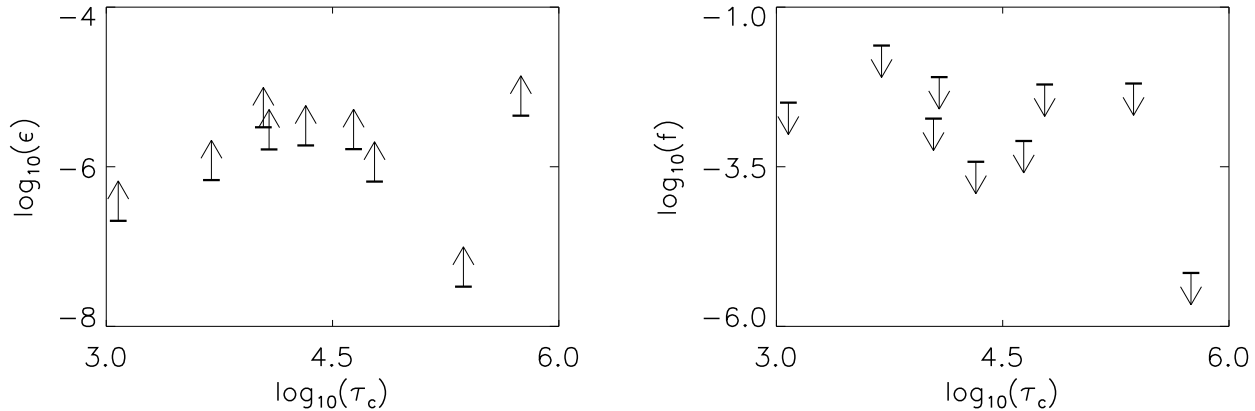}
\caption{
{\em Left panel.}
Minimum fraction of pinned vortices as a proportion of the total,
$\epsilon\leq (\Delta\nu/\nu)_{\rm max}$,
plotted versus spin-down age, measured in kyr.
{\em Right panel.}
Maximum fraction of pinned vortices that unpin thermally 
every time step in the coherent noise model,
$f\geq(\Delta\nu/\nu)_{\rm min}/(\Delta\nu/\nu)_{\rm max}$,
plotted versus spin-down age, measured in kyr.
}
\label{fig:coh9}
\end{figure*}

Figure \ref{fig:coh9} plots the lower and upper bounds implied by the data on $\epsilon$ (left panel) and $f$ (right panel) respectively against spin-down age $\tau_{\rm c}$ for the nine pulsars in Table \ref{tab:coh1}.  There is no obvious trend, although a real trend can easily be concealed, if the bounds lie far from the true physical values for the resolution-related reasons outlined in \S\ref{sec:coh3c}.  In particular, one expects $f$ to decrease steeply as $\tau_{\rm c}$ increases. As a pulsar ages, it cools, and the thermal unpinning rate should fall exponentially.  On the other hand, $F_0$ also decreases with $\tau_{\rm c}$ (Figure\ \ref{fig:coh8}), compensating for cooling at least partly.

We can estimate $f$ crudely from the general Arrhenius formula for thermally activated processes and compare with the upper bounds in the right panel of Figure\ \ref{fig:coh9}. Let us take $f \approx \exp[-E_{\rm p} ( 1 - \omega/\omega_{\rm cr}) / k_{\rm B} T]$, where $\omega$ is the vortex-superfluid angular velocity lag, and $\omega_{\rm cr}$ is the threshold lag for unpinning \citep{jon91,alp96,jah06}. For $E_{\rm p} \approx 2\,{\rm MeV}$ and $T\approx 10^6\,{\rm K}$, we obtain $E_{\rm p}/k_{\rm B} T \approx 2\times 10^4$. This implies a creep rate $f$ which is certainly consistent with, but far below, the upper bounds $f\lesssim 10^{-2}$ in the right panel of Figure\ \ref{fig:coh9}. To approach $f\sim 10^{-2}$, one requires either $| F_0 - \Delta | \sim 10^{-4} F_0$ ($f$ is dominated by the shallowest pinning sites) or  $1 - \omega/\omega_{\rm cr} \sim 10^{-4}$ (most of the stored differential rotation persists after a glitch). The upper bounds approximate the true value closely, if the experimental resolution is much better than $(\Delta\nu/\nu)_{\rm min}$ [and $\epsilon\approx (\Delta\nu/\nu)_{\rm max}$], because smaller glitches occur more commonly than larger glitches  and are therefore certain to be seen if the experimental resolution allows.  For example, in PSR J1740$-$3015, \citet{jan06} simulated microglitch detection in a noisy signal and estimated the resolution for that object to be $1\times 10^{-11}$, well below $(\Delta\nu/\nu)_{\rm min} = 7\times 10^{-10}$.  Microglitch detection simulations are needed for other objects. 
\footnote{
\citet{alp06} argued that anomalous braking indices observed
in several pulsars are partly attributable to unobserved microglitches.
\citet{jah06} related the creep rate to the time-averaged $\dot{\nu}$.
}

It should be noted that vortex creep and inhomogeneous unpinning are not independent processes.  In fact, a continuous-time coherent noise model, in which vortices unpin thermally at a continuous rate given by the Arrhenius formula, has the potential to include both phenomena self-consistently as opposite extremes of the unpinning dynamics.  That the current model does not model thermal unpinning explicitly is a direct result of it being discrete in time.

The data easily accommodate $\epsilon \gg (\Delta\nu/\nu)_{\rm max}$, because $\int_s^\epsilon ds'\,h(s')$ is typically small for $s \geq  (\Delta\nu/\nu)_{\rm max}$. However, it is debatable whether this is actually necessary. On the one hand, \citet{lyn00} measured $\epsilon = 0.017\pm0.002$, well above the largest glitch ever observed in any pulsar [$(\Delta\nu/\nu)_{\rm max}=2\times 10^{-4}$; see \citet{mel08}]. One can also argue for $\epsilon \approx 1$ on physical grounds in the context of the coherent noise model, where lattice sites and defects are microscopically separated and much more numerous than vortices. On the other hand, the aggregate value of $\epsilon$ measured by \citet{lyn00} effectively averages together different pulsars, binned over semi-decades in $\dot{\nu}$. While this approach reduces the formal error bar, it obscures the physical interpretation, given the likelihood that $\epsilon$ differs from pulsar to pulsar, and that $\langle\Delta\nu\rangle$ is dominated by the largest, rarest, and hence unobserved (over $40\,{\rm yr}$ of monitoring) glitches; see the detailed discussion in \S6 of \citet{mel08}. It is currently possible to measure $\epsilon$ reliably in only one object, PSR J0358$+$5413, which has $\epsilon \leq 7\times 10^{-5}$ \citep{mel08}.

\subsection{Aftershocks: an observational test of homogeneity
 \label{sec:coh4c}}
To the eye, the glitch statistics in Figures \ref{fig:coh1} and \ref{fig:coh2} are indistinguishable from those produced by a nearest-neighbor avalanche process, e.g.\ Figure\ 1 in \citet{war08}. Of course, the underlying collective behavior is very different: in coherent noise, the spatial correlation function $G_{\rm c}(i,j) = \langle F_{\rm p}^{(i)}F_{\rm p}^{(j)} \rangle  - \langle F_{\rm p}^{(i)} \rangle^2$  is independent of vortex separation $|{\bf x}_i-{\bf x}_j|$, whereas a self-organized critical system with nearest-neighbor avalanches exhibits long-range correlations on all scales, with $G_{\rm c} \propto |{\bf x}_i-{\bf x}_j|^{-\beta}$ and $\beta > 0$. Unfortunately, we cannot measure $G_{\rm c}$ directly in a neutron star.  However, it turns out that we can still discriminate between homogeneous (e.g.\ coherent noise) and inhomogeneous (e.g.\ avalanches) glitch mechanisms observationally, by taking advantage of temporal correlations embedded in the data. Below, we propose three related observational tests of this kind, all of which are practical in the medium term.
\begin{enumerate}
\item
{\em Aftershocks.}
Aftershocks occur in a coherent noise process because, 
after a large glitch with $F_{\rm M} \geq F_0+\Delta$,
the threshold distribution $g(F_{\rm p})$ is repopulated evenly;
that is, $\langle F_{\rm p} \rangle$ immediately after a large glitch
is less than $\langle F_{\rm p} \rangle$ for the time-averaged $g(F_{\rm p})$,
which is depleted at low $F_{\rm p}$.
Consequently, more vortices than usual unpin during the next few time steps
\citep{sne97}.
In contrast, aftershocks are not produced by nearest-neighbor avalanches,
because successive glitches arise from the relaxation of 
insular capacitive domains and are therefore independent.  \footnote{As defined in \citet[][and references therein]{alp96}, capacitive domains may indeed be connected, and hence their relaxation not independent.  In the wider (e.g. experimental) SOC literature, capacitive domains
are defined such that all connected "subdomains" constitute one
domain. See for example \cite{sor91,jen98}.
 Defined this way, the relaxation of successive domains is
statistically independent (except for system-wide avalanches),
as verified by experiments with sand and rice grains
\citep{ros93}.}

Aftershocks cannot be analyzed by the time-averaged, mean-field, analytic theory
in \S\ref{sec:coh2c},
which contains no information about temporal correlations.
\footnote{
The time-averaged theory assumes a unique glitch size
$s(F_{\rm M})$ for every Magnus force $F_{\rm M}$,
whereas in reality the automaton in \S\ref{sec:coh2b} produces a
power-law distribution of sizes for fixed $F_{\rm M}$;
see Figure\ 8 of \citet{sne97}.
}
But Monte-Carlo simulations of the automaton in \S\ref{sec:coh2b}
reveal the aftershock effect clearly.
Figure \ref{fig:coh10} displays the conditional size probability density function
$h(s,t+\Delta t | s',t)$ for $s'\geq\epsilon$, $0.67\epsilon$, $0.5\epsilon$,
and $0.33\epsilon$ (color coded from dark to light)
and compares it with the full size distribution 
$h(s)=\sum_{s'} h(s,t+\Delta t | s',t)$
(lightest grey).
Clearly,
there is a relative excess of large glitches following a large glitch,
especially for $s'=\epsilon$.
Note that the aftershocks discussed here are standard glitches;
they are not the same as the time-resolved secondary spin up events
noted in the Crab by \citet{won01}, which occur 20--40\,d after a glitch.

\begin{figure}[ht!]
\epsscale{0.7}
\begin{center}
\plotone{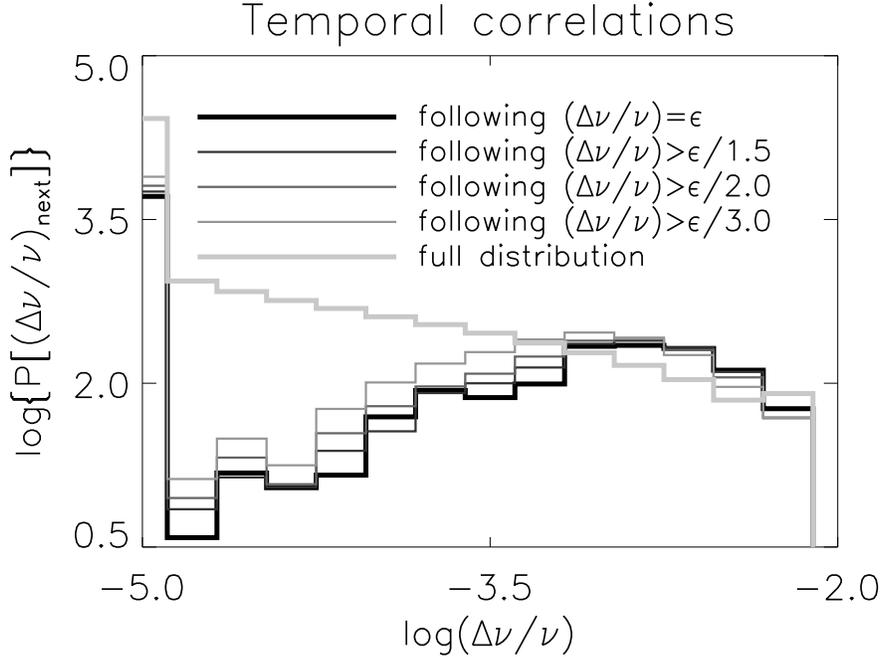}
\end{center}
\caption{
Conditional size probability density function $h(s,t+\Delta t | s',t)$
for $s'\geq\epsilon$ (thick black histogram), $0.67\epsilon$ (dark grey), 
$0.5\epsilon$ (medium grey),
and $0.33\epsilon$ (light grey),
together with the full probability density function
$h(s)=\sum_{s'} h(s,t+\Delta t | s',t)$ (lightest grey).
The conditional function describes the probability of getting a glitch
of size $s$ at time $t+\Delta t$ following a glitch of size $s'$
at time $t$.
The excess of large glitches following a large glitch is a clear signature
of the aftershock effect.
}
\label{fig:coh10}
\end{figure}

\item
{\em Glitch lifetimes.}
The distribution of glitch lifetimes, or rise times (i.e.\ durations; cf.\ waiting times), is also a power law in a coherent noise process, with exponent $\approx 1.0$ (plus a small logarithmic correction) \citep{sne97}.  The power law stems from the memory effect described in \S\ref{sec:coh3a}.  Its exponent is the same for all forms of $\psi(F_{\rm M})$.  In contrast, an avalanche process leads to a range of exponents for the lifetime power law \citep{war08}.  Testing this prediction is ambitious.  At present, the best measurement of glitch durations is an upper limit of $\approx 40\,{\rm s}$, which comes from continuous monitoring of Vela \citep{dod02}. However, it may be possible to improve on the resolution currently available through single pulse timing and daily monitoring with low-radio-frequency arrays currently under construction [M. Bailes and R. Bhat, private communication, \citet{dod08}].

It should be noted, however, that a single component model like the one presented here cannot accurately describe glitch lifetimes, which depend on the coupling between the stellar interior and its crust.  This coupling is thought to rely on the presence of superfluid neutrons and superconducting protons in the core of the neutron star \citep{alp06}.  A two component coherent noise model, capable of addressing the issue of crust-core coupling, will be the subject of a future paper.

\item
{\em Size versus waiting time.}
The coherent noise model predicts a statistical correlation between the size of a glitch and the time since the previous glitch.  In contrast, an avalanche process exhibits no such correlation.  For example, it can store Magnus stress in metastable reservoirs over a long period of time, punctuated by minor glitches, until a major glitch releases most of the stress all at once \citep{jen98,sor04,mel08,war08}.  The necessary homogeneity of the model presented here does not allow for the unique treatment of different components of the stellar interior.  If, as is discussed in detail in \cite{alp06}, we could differentiate between the capacitive and resistive regions of the superfluid, we would expect similar correlations between the pulsar spin-down rate, the interglitch waiting time and glitch size.  In particular, by identifying regions of the superfluid where fluctuations in vortex pinning arise thermally and regions where vortices are always pinned, the effective pinned fraction $\epsilon$ becomes spatially dependent \citep{alp06}.  

The coherent noise correlation is imperfect because of the memory effect described in \S\ref{sec:coh3a}; two events with the same $F_{\rm M}$ can lead to different $s$ values, according to whether or not $g(F_{\rm p})$ is depleted at low $s$ by a prior sequence of relatively small $F_{\rm M}$.  Nevertheless, Figure\ \ref{fig:coh11} demonstrates that it is present on average.  In the upper panel of Figure\ \ref{fig:coh11}, the size of a glitch is plotted against the time since the previous glitch (normalized to $\lambda^{-1}$) for a representative Monte-Carlo simulation of the coherent noise model.  The associated linear Pearson correlation coefficient is $r=0.57$ for $10^6$ events, implying a substantial correlation.  In the lower panel of Figure\ \ref{fig:coh11}, the same plot is constructed for the 23 glitches of PSR J0534$+$2200 (black asterisks).  The associated Pearson coefficient is $r=-0.22$, implying a weaker correlation.  Likewise, we find $r=0.16$ for PSR J0835$-$4510, which is not plotted.  We advise caution when interpreting these results.  (i) For small $N_{\rm g}$, a sample of $N_{\rm g}$ events does not exhibit the correlation in the upper panel at a statistically significant level.  (ii) The correlation stems from the aftershock effect.  It is dominated by the small-$s$ events populating the dark band in the upper panel, the part of the distribution that is measured least reliably (e.g.\ unobserved microglitches).

Many authors have previously searched empirically for a size versus waiting time correlation in glitch data and found none, e.g.\ \citet{wan00} and \citet{won01}.  However, all previous studies have analyzed the statistics from all glitching pulsars in aggregate (grey asterisks in the lower panel of Figure\ \ref{fig:coh11}), in an effort to maximize the number of data points.  Since pulsars have different $\lambda$ and $\langle \Delta \nu \rangle$ \citep{mel08}, both axes of the potential correlation are washed out in aggregate data.  It is therefore vital to test pulsars individually.  In the case of the much-studied Vela pulsar, models that include correlations between the spin-down rate and the interglitch waiting time have successfully described glitch behaviour \citep[e.g.][]{alp84}.  Similar analyses have been conducted by \cite{mar04} and \cite{mid06} on PSR J0537--6910, resulting in claims that the interglitch interval can be predicted to within a few days.

\begin{figure}[ht!]
\epsscale{1.0}
\begin{center}
\plotone{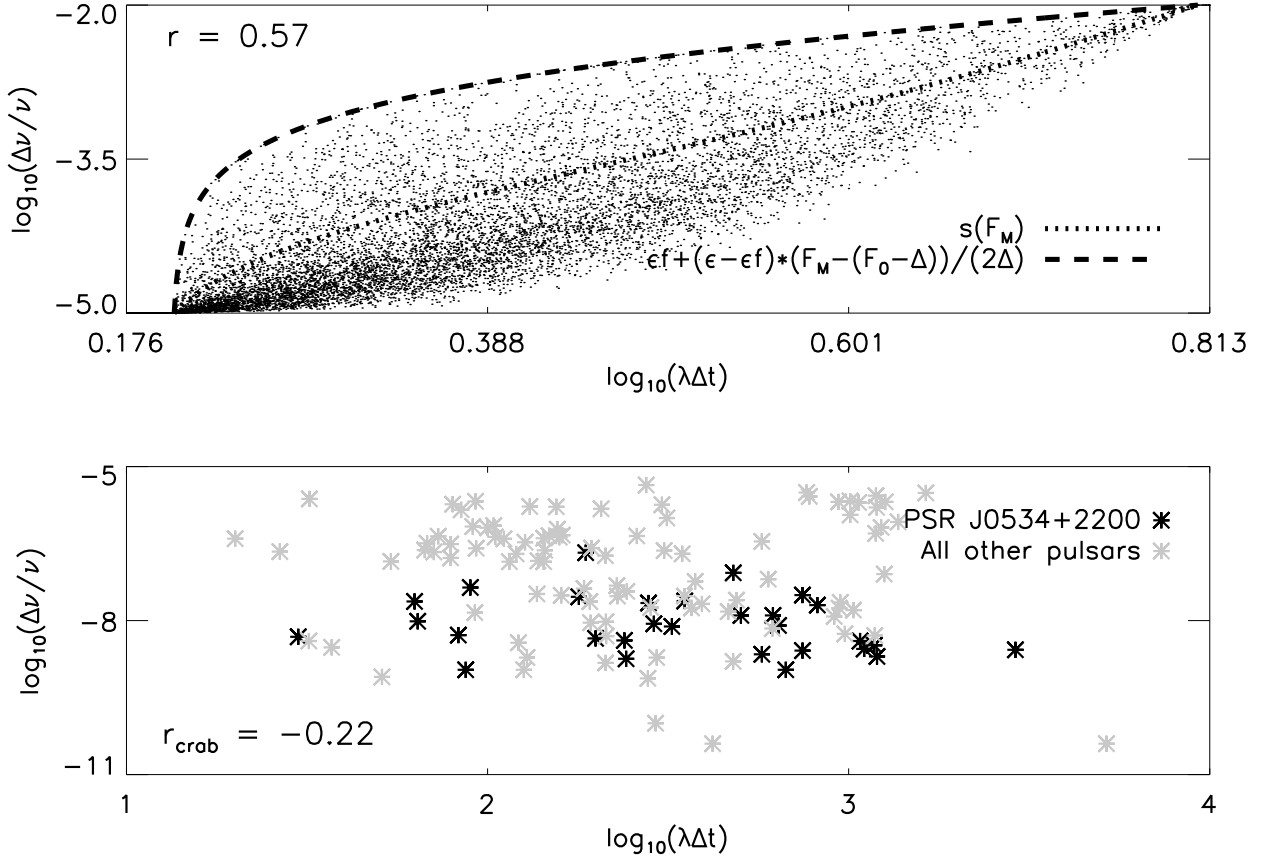}
\end{center}
\caption{
{\em Upper panel.}
Glitch size, $s$, versus the normalized time since the previous glitch, $\lambda\Delta t$,
for a representative simulation of the coherent noise model,
with 
$N=10^6$, $\epsilon=10^{-2}$, $f=10^{-3}$, $F_0=4.0\sigma$, and $\Delta=0.6F_0$.
Each dot marks one glitch.
The distribution is bounded above by the function 
$\epsilon f+\epsilon (1- f)[ \lambda \sigma \Delta t-(F_{0}-\Delta)] / (2\Delta)$ 
(dashed curve)
and is centered around the time-averaged relation
$s(F_{\rm M}=\lambda /\sigma\Delta t)$ from equation (\ref{eq:coh6})
(dotted curve).
The Pearson linear correlation coefficient $r$ is printed on the plot.
{\em Lower panel.}
Glitch size, $s$, versus the normalized time since the previous glitch, $\lambda\Delta t$,
for PSR J0534$+$2200 (black asterisks)
and for every glitch observed to date (grey asterisks).
}
\label{fig:coh11}
\end{figure}

\end{enumerate}

For the definition of $\sigma$ that assumes that the Magnus stress is proportional to the time elapsed ($\sigma_1$), the coherent noise model analyzed in this paper is incomplete in one major respect. It does not predict endogenously the waiting-time distribution and hence the mean rate $\lambda$. Rather, $\lambda$ is measured observationally and put into the model by hand through $\psi(F_{\rm M})$. There is nothing wrong with this approach, of course; it takes advantage of a well-determined observational fact to construct the theory. But it does mean that the model is incomplete.  In particular, as it stands, the model cannot answer important questions like why some pulsars glitch and others do not, and why, of the objects that do glitch, some are more active than others. Moreover, it can only be applied to objects where enough glitches have been observed to measure $\lambda$ reliably. On the positive side of the ledger, there is no reason in principle why a microphysical theory of the waiting-time distribution and hence $\lambda$ cannot be developed by studying the temporal behavior of single-vortex unpinning triggers like thermal activation \citep{alp84,jon91,jah06}. Once available, such a theory, combined with the coherent noise process to explain {\em collective} unpinning, and generalized to include the radial dependence of $\phi(F_{\rm p})$, may offer a path to a complete glitch theory.  The beginnings of such an approach are contained in Equations (\ref{eq:sig2}) and (\ref{eq:domega}) in \S\,\ref{sec:coh4a}.  A similar approach may also prove fruitful in interpreting classic experiments on magnetic flux creep in type II superconductors \citep{fie95,bas98}. 

\subsection{Observational verification}
\label{sec:obs}

\begin{figure}[ht!]
\epsscale{1.}
\begin{center}
\plotone{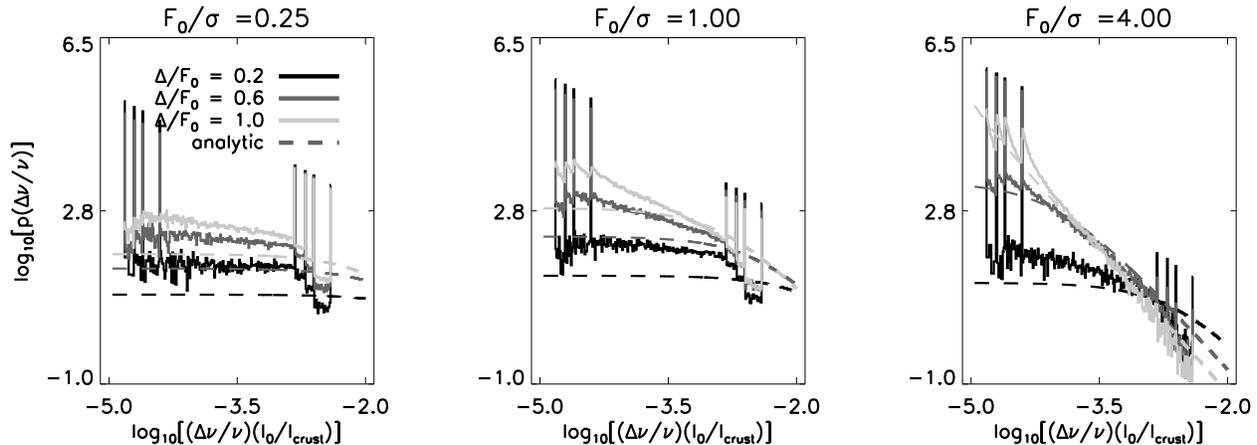}
\end{center}
\caption{
Theoretical size and pinning threshold distributions as functions of the pinning parameters $F_0$ and $\Delta$.  Parameters are the same as for Figure \ref{fig:coh2}, but it is now assumed that unpinned vortices traverse at random one of four regions moments of inertia accounting for 0.15, 0.2, 0.25 and 0.4 of the total before repinning.  The fractional glitch size is also multiplied by the ratio of the total stellar moment of inertia to the moment of inertia of the crust ($I_0/I_{\rm{c}}$).
}
\label{fig:coh12}
\end{figure}

The coherent noise model in this paper is unrealistic in three important respects, which render it difficult to verify observationally. (i) The model is discrete, not continuous, in time. (ii) It is homogeneous, so it cannot include the radial dependence of pinning strength (or creep rate), and the moments of inertia of the regions from/through/to which the vortices move, in a self-consistent way. For example, vortex motion per se is not modelled; we do not track where individual vortices go when they unpin. (iii) It allows no feedback of the vortex motion on the two stellar components (pinned superfluid and crust), and hence does not determine the observable rotational dynamics in an internally self-consistent way. These weaknesses do not interfere with the main physical result: namely, and counterintuitively, that you get a power-law distribution of event sizes from a Poisson distribution of waiting times, {\em in spite of the system being homogeneous, and without nearest-neighbor avalanches occurring}.  These three weaknesses significantly reduce the ease with which this model can be distinguished from other glitch models, such as those invoking avalanche dynamics \cite[][for example]{war08}.  The appearance of features such as aftershocks, correlations between glitch sizes and waiting times, and particular power-law exponents, may depend strongly on the coupling of the different stellar components, and hence require that the model account for this coupling.

In a first, crude attempt to generalize the coherent noise model for multiple components, we augment the model automaton described in \S\,\ref{sec:coh2b} so that the vortices unpinned during each glitch to pass randomly through one of four regions of the star before repinning.  The choice of four regions is arbitrary and merely illustrative.  Following the vortex creep model presented in \cite{alp86}, we assume that the size of a glitch depends on both the number of vortices that unpin, and the fraction of the total stellar moment of inertia through which the unpinned vortices move before repinning.  A homogeneous model does not record the starting and end points of the vortex motion during a glitch, and hence cannot accurately predict the change in angular momentum of the superfluid resulting from the (un)pinning process.  As a worked example, we assume that the fractional volume traversed by unpinned vortices falls into one of four broad bins, accounting for 0.15, 0.2, 0.25, and 0.4 of the total moment of inertia of the star $I_0$.  The glitch size is then calculated as the product of the unpinned vortex fraction and traversed moment-of-inertia fraction.  The results of Monte-Carlo simulations using the modified ``multi-component'' automaton are shown in Figure \ref{fig:coh12}.  The main effect, when compared to the single component model shown in Figure \ref{fig:coh2} for the same values of $F_0/\sigma$, $\Delta /\sigma$, $\epsilon$, and $f$, is to blur the upper and lower edges of the power law (without changing much the exponent in between).  That is to say, multiple components change the effective value of $\epsilon$.  

We note that in the realistic case where all unpinned vortices traverse similar volumes, the multiple spikes appearing in the size distributions in Figure \ref{fig:coh12} merge into single peaks at both ends of the distribution.  A self-consistent study of this important effect awaits future work.

In closing, we emphasize again that the model in this paper is highly idealized.  Quantitative conclusions drawn from fitting the model to observational data should be viewed cautiously until more glitches have been observed. 

\acknowledgments 
The authors thank Stuart Wyithe for expert advice 
on how to rigorously fit the theory to the data in \S\ref{sec:coh3}.  We thank the referee, Dr Ali Alpar, for his thorough reading of the paper; his insightful comments have greatly improved this manuscript.  This research was supported by an Australian Postgraduate Award,
the University of Melbourne Research Grant Scheme,
and the University of Melbourne-CSIRO Collaborative Research Support Scheme.

\bibliographystyle{apj}
\bibliography{glitchstat}

\appendix
\section{Analytic results for a top-hat distribution 
 of pinning thresholds
 \label{sec:cohappa}}
In this appendix, we write down explicit analytic formulas for the 
main results of the mean-field theory when applied to
a top-hat distribution of pinning thresholds,
given by (\ref{eq:coh1}),
and an exponential distribution of Magnus stresses,
given by (\ref{eq:coh2}).
To assist readability,
we redefine the force (per unit length) variables 
$F_{\rm M}$, $F_{\rm p}$, $F_0$, and $\Delta$,
appearing in (\ref{eq:cohappa0b})--(\ref{eq:cohappa5}) below,
to be normalized versions of their counterparts in the main text,
measured in units of $\sigma$.
We also introduce the auxiliary function
\begin{equation}
 \lambda(x) = 1 - f + f\exp(x)
\label{eq:cohappa0a}
\end{equation}
and the parameter 
\begin{equation}
 \mu = \mu(f,F_0,\Delta) =
 \ln\left[
  \frac{\lambda(F_0+\Delta) }
   {\lambda(F_0-\Delta) }
 \right]~.
\label{eq:cohappa0b}
\end{equation}

\begin{enumerate}
\item
Time-averaged threshold distribution at occupied pinning sites:
\begin{eqnarray}
 g(F_{\rm p})
 & = &
 f \exp(F_{\rm p}) [ \mu \lambda(F_{\rm p}) ]^{-1}
 \nonumber \\
 & & 
 \quad \times
 H(F_{\rm p} - F_0 + \Delta)
 H(-F_{\rm p} + F_0 + \Delta)~.
\label{eq:cohappa1}
\end{eqnarray}
\item
Glitch size as a function of Magnus stress:
\begin{eqnarray}
 s(F_{\rm M})
 & = &
 \epsilon f + \epsilon(1-f) \mu^{-1}
 \ln\left[
  \frac{\lambda(F_{\rm M}) }
   {\lambda(F_0-\Delta) }
 \right]
 \nonumber \\
 & & 
 \quad \times
 H(F_{\rm M} - F_0 + \Delta)
 H(-F_{\rm M} + F_0 + \Delta) 
 \nonumber \\
 & &
 +
 \epsilon f H(- F_{\rm M} + F_0 - \Delta)
 \nonumber \\
 & &
 +
 \epsilon H(F_{\rm M} - F_0 - \Delta)~.
\label{eq:cohappa2}
\end{eqnarray}
\item
Time-averaged probability distribution function of glitch sizes:
\begin{eqnarray}
 h(s)
 & = &
 \frac{ f \mu \lambda(F_0-\Delta) }
  { \epsilon (1-f) }
 \exp\left[
  \frac{\mu(s-\epsilon f)}{\epsilon(1-f)}
 \right]
 \nonumber \\
 & &
 \quad \times
 \left\{
  -(1-f) +
  \lambda(F_0-\Delta)
  \exp\left[
   \frac{\mu(s-\epsilon f)}{\epsilon(1-f)}
  \right]
 \right\}^{-2}
 \nonumber \\
 & & 
 \quad \times
 H(s-\epsilon f)
 H(-s + \epsilon)
 \nonumber \\
 & &
 +
 [ 1 - \exp(-F_0+\Delta) ]
 \delta(s-\epsilon f)
 \nonumber \\
 & &
 + 
 \exp(-F_0-\Delta)
 \delta(s-\epsilon)~.
\label{eq:cohappa3}
\end{eqnarray}
\end{enumerate}

If $f$ is small, such that the interval $f \leq s/\epsilon \leq 1$ spans several decades,
$h(s)$ tends to a universal power law, namely $h(s) \propto s^{-2}$,
in the regime $s\gg \epsilon f$.
To see this mathematically, consider the limit where $f$ is small enough
such that $\mu$ is also small.
A sufficient condition for this ordering to obtain is 
\begin{equation}
 f\exp(F_0+\Delta) \ll 1~.
\label{eq:cohappa4}
\end{equation}
When (\ref{eq:cohappa4}) is satisfied, we find
$\mu \approx f [\exp(F_0+\Delta) -\exp(F_0-\Delta) ]$
and hence
\begin{equation}
 h(s) \approx 
 \frac{\epsilon \exp(-F_0+\Delta)}{\exp(2\Delta)-1}
 \left[
 s +
 \frac{\epsilon}{\exp(2\Delta)-1}
 \right]^{-2}
\label{eq:cohappa5}
\end{equation}
for $\epsilon f \ll s < \epsilon$.
A similar result follows for any other physically reasonable choices of 
$\phi(F_{\rm p})$ and $\psi(F_{\rm M})$
\citep{sne97}.

\section{Maximum likelihood algorithm for fitting $h(s)$
 \label{sec:cohappb}}
In this appendix, we describe briefly the fitting algorithm used to
produce Figures \ref{fig:coh4} and \ref{fig:coh6} and Table \ref{tab:coh1},
to help the reader reproduce the results.
The algorithm can handle relatively small data sets meaningfully, 
but of course the statistical significance of its output
improves as $N_{\rm g}$ increases.

\begin{figure*}
\epsscale{1.0}
\plotone{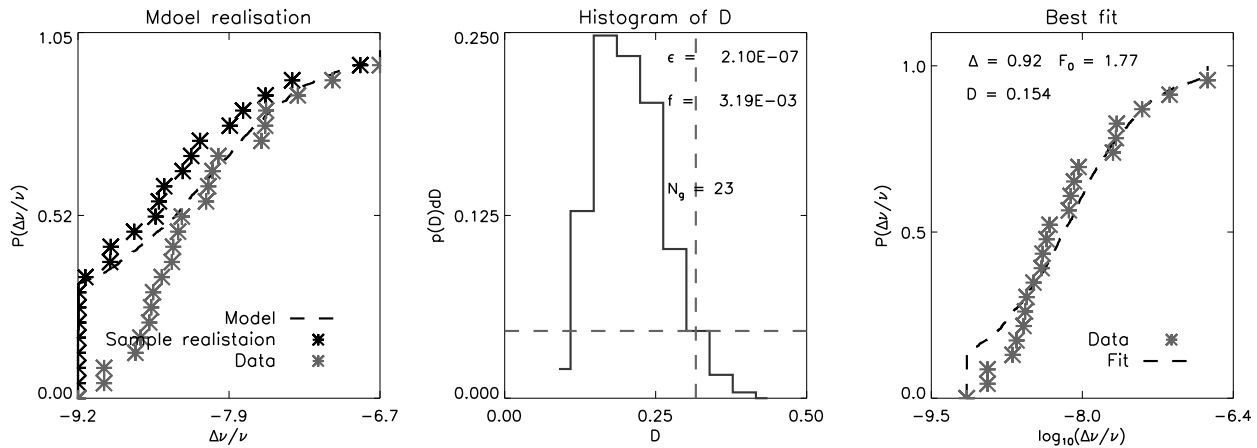}
\caption{
Graphical demonstration of the maximum likelihood algorithm for fitting $h(s)$.
{\em Left panel.}
A model is chosen by selecting $F_0$ and $\Delta$.
Many ($\sim 10^3$) realizations of the model are created by sampling the
theoretical distribution $N_{\rm g}$ times.
One representative realization is plotted here (black asterisks),
together with the continuous cumulative probability distribution
from which it is drawn (dashed curve)
and the observational data (grey asterisks).
{\em Middle panel.}
The Kolmogorov-Smirnov $D$ statistic quantifies
the separation between the realization and the underlying continuous distribution.
A frequency histogram of the $D$ statistic is constructed from all the realizations
(solid staircase),
normalized to unit area.
The separation between the data and the continuous distribution, $D_{\rm data}$,
is also computed.
The relative likelihood of a trial pair $(F_0,\Delta)$, denoted by ${\cal L}$,
is defined as the height of the histogram at $D=D_{\rm data}$,
where the dashed horizontal and vertical lines meet.
{\em Right panel.}
The foregoing procedure is repeated for many combinations of $F_0$ and $\Delta$,
each time yielding a $D$ histogram, a value of $D_{\rm data}$, 
and a relative likelihood ${\cal L}$.
The best fit parameters maximize ${\cal L}$.
Here, the best fit is achieved for
$\epsilon=2.1\times 10^{-7}$, $f=3.2\times 10^{-3}$, $F_0/\sigma=1.8$, 
and $\Delta/F_0 = 0.92$.
The associated cumulative probability distribution (dashed curve)
is plotted over the observational data (grey asterisks).
}
\label{fig:coh3}
\end{figure*}

Figure \ref{fig:coh3} demonstrates how to obtain a best fit
in three steps,
taking as an example the Crab (PSR J0534$+$2200),
which has glitched $N_{\rm g}=23$ times.
(i)
Starting with a particular choice of $F_0/\sigma$ and $\Delta/F_0$,
we create many ($\sim 10^3$) realizations of the model by sampling the 
continuous,
theoretical $h(s)$ $N_{\rm g}$ times to construct each realization.
A representative realization is plotted as a cumulative probability distribution 
in the left panel of Figure\ \ref{fig:coh3}
(black asterisks), 
together with the continuous distribution
$\int_{\epsilon f}^s ds' \, h(s')$ 
from which the realization is sampled (dashed curve),
and the observational data (grey asterisks).
(ii) 
We compute the maximum unsigned separation $D$
between each realization and the continuous distribution,
i.e.\ the Kolmogorov-Smirnov statistic,
\footnote{
By tailoring $\epsilon f$ to match $(\Delta\nu/\nu)_{\rm min}$,
we artificially impose the restriction $D\geq N_{\rm g}^{-1}$,
caused by the guaranteed discrepancy in the leftmost bin.
Although the restriction varies from pulsar to pulsar,
it applies equally to all the model realizations and the
observational data in any individual object,
so its distorting influence is mild.
}
and bin the results to produce a frequency histogram of $D$ which is
unweighted by bin width and normalized to have unit area,
as in the middle panel of Figure\ \ref{fig:coh3}.
(Note that the plotted histogram is representative;
it does not correspond to the choice of $F_0$ and $\Delta$
that gives the best fit.)
We then compute the maximum unsigned separation $D_{\rm data}$
between the observational data and the continuous distribution.
The height of the histogram at $D=D_{\rm data}$,
where the dashed lines intersect in the middle panel of Figure\ \ref{fig:coh3},
is called the relative likelihood ${\cal L}(F_0/\sigma, \Delta/F_0)$
for that particular choice of $F_0/\sigma$ and $\Delta/F_0$.
\footnote{
The relative likelihood ${\cal L}$ only has meaning when used to compare
models with different $(F_0/\sigma,\Delta/F_0)$ in the same pulsar.
It cannot be used to compare models across different pulsars.
${\cal L}$ is neither a probability nor a probability density.
If we have ${\cal L}(F_0,D) = 1.5 {\cal L}(F'_0, D')$,
say, we can conclude that model
$(F_0, D)$ is more likely than model $(F'_0, D')$,
but not that it is 1.5 times more likely.
}
(iii) We repeat for a range of $F_0/\sigma$ and $\Delta/F_0$.
The best fit parameters are those that maximize ${\cal L}(F_0/\sigma, \Delta/F_0)$.
For example, in Figure\ \ref{fig:coh3}, the maximum ${\cal L} = 0.23$ is achieved
for $(F_0/\sigma,\Delta/F_0)=(1.8,0.92)$
(here, the gridding differs slightly from Figure\ \ref{fig:coh4}).
The corresponding continuous cumulative distribution
is plotted as a dashed curve in the right panel of Figure\ \ref{fig:coh3},
together with the observational data (grey asterisks).

\end{document}